\global\def\draftcontrol{0}

   \def\versionno{ Holographic Renormalization -- draft   }
\catcode`\@=11

\expandafter\ifx\csname draftcontrol\endcsname\relax\global\def\draftcontrol{0}
\fi

{\count255=\time\divide\count255 by 60
\xdef\hourmin{\number\count255}
\multiply\count255 by-60\advance\count255 by\time
\xdef\hourmin{\hourmin:\ifnum\count255<10 0\fi\the\count255}}
\def\draftdate{\number\month/\number\day/\number\year\ \ \ \hourmin }

\newcommand\makepapertitle{\par
  \begingroup
    \renewcommand\thefootnote{\@fnsymbol\c@footnote}%
    \def\@makefnmark{\rlap{\@textsuperscript{\normalfont\@thefnmark}}}%
    \long\def\@makefntext##1{\parindent 1em\noindent
            \hb@xt@1.8em{%
                \hss\@textsuperscript{\normalfont\@thefnmark}}##1}%
     \newpage
     \global\@topnum\z@   % Prevents figures from going at top of page.
     \@makepapertitle
     \thispagestyle{empty}\@thanks
  \endgroup
  \setcounter{footnote}{0}%
  \global\let\thanks\relax
  \global\let\makepapertitle\relax
  \global\let\@makepapertitle\relax
  \global\let\@thanks\@empty
  \global\let\@author\@empty
  \global\let\@date\@empty
  \global\let\@title\@empty
  \global\let\title\relax
  \global\let\author\relax
  \global\let\date\relax
  \global\let\and\relax
  \def\version{\let\version\@version\@gobble}
}
\def\@makepapertitle{%
  \newpage
   \ifnum\draftcontrol=1 {}
   \version\versionno
   \vskip 3em%
   \else
   \hfill\hbox to 3cm {\parbox{4cm}{\@pubnum}\hss}%
   \vskip 3em%
   \fi
   \begin{center}%
   \let \footnote \thanks
     {\LARGE {\@title}}%
     \vskip 1.5em%
     {\normalsize%\large
       \lineskip .5em%
       \begin{tabular}[t]{c}%
         \@author
       \end{tabular}\par}%
     \vskip 1.5em%
     {\@bstract}%
     \end{center}%
     \vskip 1.5em
     \@date%
   \par
}

\gdef\@pubnum{}
\def\pubnum#1{%
  \gdef\@pubnum{#1}}

\gdef\@bstract{}
\def\Abstract#1{%
  \gdef\@bstract{%
   \parbox{\textwidth-0pc}{%
   \centerline{\bf Abstract}\penalty1000%
\kern.2cm%
\noindent%\abstractfont \baselineskip=12pt
\renewcommand\baselinestretch{1.0}%
{#1}}}
}

\def\ps@paper{\let\@mkboth\@gobbletwo%
     \ifnum\draftcontrol=1
    \def\@oddfoot{\hbox to \textwidth{\tiny \versionno \hfil\tiny\draftdate}%
    \hskip -\textwidth \hbox to \textwidth{\hfil\rm\thepage\hfil}}%
     \else\def\@oddfoot{\hbox to \textwidth{\hfil\rm\thepage\hfil}}
     \fi
     \let\@evenfoot\@oddfoot
}
\def\body{\clearpage
          \pagestyle{paper}
    }
\def\@version#1{\ifnum\draftcontrol=1
\typeout{}\typeout{#1}\typeout{}
\vskip3mm\centerline{\hbox{\fbox{\normalsize{\tt DRAFT -- #1 -- }
                   {\draftdate}}}}\vskip3mm
\fi}
\let\version\@version
\long\def\eqlabel#1{\ifnum\draftcontrol=1
                    \tag@false  % there are some problems with multline without this
                    \tag*{(\theequation) \hbox to -0.2cm{\hspace{0cm}\small{#1}\hss}}
                    \refstepcounter{equation}
                    \edef\@currentlabel{\theequation}
                    \ltx@label{#1}          % use old LaTeX \label instead of new definition
                    \else
                    \label{#1}
                    \fi
                    }
\let\st@bibitem\@bibitem
\let\st@lbibitem\@lbibitem
\ifnum\draftcontrol=1
  \def\@bibitem#1{%
    \st@bibitem{#1}\a@@label{#1}\ignorespaces}
  \def\@lbibitem[#1]#2{%
    \st@lbibitem[#1]{#2}\a@@label{#2}\ignorespaces}
  \def\a@@label#1{%
    \gdef\a@lab{\smash{\normalfont\small#1}}
    \ifvmode
      \if@inlabel
        \global\setbox\@labels\hbox{%
          \llap{\a@lab\let\a@lab\relax
                \kern\@totalleftmargin\kern\marginparsep}%
          \box\@labels}%
      \fi
    \fi}
\fi
\documentclass[11pt,letterpaper]{article}
\usepackage{amsmath,amssymb,array,calc,epsfig}
\ifnum\draftcontrol=1
\fi
\renewcommand\baselinestretch{1.25}
\renewcommand\section{\@startsection {section}{1}{\z@}%
                                   {-3.5ex \@plus -1ex \@minus -.2ex}%
                                   {2.3ex \@plus.2ex}%
                                   {\normalfont\large\bfseries}}
\renewcommand\subsection{\@startsection{subsection}{2}{\z@}%
                                   {-3.25ex\@plus -1ex \@minus -.2ex}%
                                   {1.5ex \@plus .2ex}%
                                   {\normalfont\normalsize\bfseries}}
\renewcommand\subsubsection{\@startsection{subsubsection}{3}{\z@}%
                                   {-3.25ex\@plus -1ex \@minus -.2ex}%
                                   {1.5ex \@plus .2ex}%
                                   {\normalfont\normalsize\it}}
\renewcommand\paragraph{\@startsection{paragraph}{4}{\z@}%
                                   {-3.25ex\@plus -1ex \@minus -.2ex}%
                                   {1.5ex \@plus .2ex}%
                                   {\normalfont\normalsize\bf}}

\numberwithin{equation}{section}

\def\revise#1       {\raisebox{-0em}{\rule{3pt}{1em}}%
                     \marginpar{\raisebox{.5em}{\vrule width3pt\
                     \vrule width0pt height 0pt depth0.5em
                     \hbox to 0cm{\hspace{0cm}{%
                     \parbox[t]{4em}{\raggedright\footnotesize{#1}}}\hss}}}}

\def\sqr#1#2{{\vcenter{\vbox{\hrule height.#2pt
 \hbox{\vrule width.#2pt height#1pt \kern#1pt
 \vrule width.#2pt}\hrule height.#2pt}}}}

\def\aa1{\phi}
\def\cc1{\psi}

\catcode`\@=12

\begin{document}

%%%
%%%%%% text starts here
%%%%%%%%%

\title{\bf A Note on Holographic Renormalization of Probe D-Branes}

\pubnum{%
arXiv:0903.xxxx}
\date{March 2009}

\author{
\scshape Paolo Benincasa\\[0.4cm]
\ttfamily Center for Particle Theory \& Department of Mathematical Sciences\\
\ttfamily Science Laboratories, South Road, Durham DH1 3LE, United Kingdom\\[0.2cm]
\small \ttfamily paolo.benincasa@durham.ac.uk
}

\Abstract{A great deal of progress has been recently made in the study of holography for non-conformal
branes. Considering the near-horizon limit of backgrounds generated by such branes, we discuss the holographic 
renormalization of probe D-branes in these geometries. More specifically, we discuss in some detail 
systems with a codimension-one defect.
%(i.e. D$p$/D$(p+2)$ systems) for which two descriptions for the embedding of the probe branes are possible. 
For this class of systems, 
the mode which describes the probe branes wrapping a maximal $S^2$ in the transverse space behaves like a 
free massive scalar propagating in a higher-dimensional (asymptotically) $AdS_{\mbox{\tiny $q+1$}}$-space. 
%In principle, the actual number of the $AdS$-space dimensions would be fractional. 
%We map the problem to probe branes wrapping and $AdS_{q+1}\times S^{2}$ space with $q$ being a
%general integer. 
The counterterms needed are then the ones of a free massive scalar in asymptotically $AdS_{q+1}$. The
original problem can be recovered by compactifying the asymptotically $AdS$-space on a torus and finally 
performing the analytic continuation of $q$ to the value of interest, which can be fractional. 
We compute the one-point correlator for the operator dual to the embedding function. 
%The standard $AdS/CFT$ prescription can be applied at the level of the higher-dimensional $AdS$-space, 
%compactifying then the extra-dimensions on a torus.
We finally comment on holographic renormalization in the more general cases of codimension-$k$ defects 
($k=0,\,1,\,2$). In all the cases the embedding function exhibits the behaviour of a free massive scalar in an
$AdS$-space and, therefore, the procedure outlined before can be straightforwardly applied. 
Our analysis completes the discussion of holographic renormalization of probe D-branes started by
Karch, O'Bannon and Skenderis.}

\makepapertitle

\body

\version\versionno

\section{Introduction}

Gauge/gravity correspondence \cite{Maldacena:1997re, Gubser:1998bc, Witten:1998qj, Aharony:1999ti}
provides a powerful tool to investigate the dynamics of strongly coupled
gauge theories. The original formulation \cite{Maldacena:1997re} conjectures the equivalence between
supergravity on the ``near-horizon'' geometry generated by a stack of $N$ coincident D$3$-branes 
($AdS_{5}\times S^5$) and the gauge theory 
(four-dimensional $\mathcal{N}=4$ $SU(N)$ Supersymmetric Yang-Mills) living on the 
boundary of $AdS_{5}$, which describes the brane modes decoupled from the bulk. 
It can be straightforwardly extended to any asymptotically $AdS\times\mathcal{M}$ geometry, 
$\mathcal{M}$ being a compact manifold. This conjectured equivalence
is made precise by identifying the string partition function with the generating function for the gauge
theory correlators, with the boundary value of the bulk modes acting as source of the correspondent
gauge theory operator.

An important issue is the presence of divergences on both sides of the correspondence: the divergences
appear in the UV-region on the gauge theory side and in the IR-region on the gravity side. These two 
divergence structures are related to each other \cite{Susskind:1998dq}, which can be intuitively 
understood by thinking that, on the gravity side, going to the IR-region means approaching the
boundary (where the gauge theory sits). Holographic renormalization consistently deals with these
IR divergences \cite{Henningson:1998gx, Henningson:1998ey, Balasubramanian:1999re, deHaro:2000xn,
Skenderis:2000in, Bianchi:2001de, Bianchi:2001kw, Skenderis:2002wp, Papadimitriou:2004ap, Papadimitriou:2004rz}
so that physical quantities can be computed. The first step is to find a solution for the bulk fields
in a neighbourhood of the boundary and regularize the action by means of an IR-regulator. By inserting
the solution previously found in the action, one can read off the (finite number of) terms which diverge once
the regulator is removed. A counterterm action can then be constructed as invariant local functional of the
metric and fields on the boundary of the regulated space-time in such a way that these terms are cancelled.

This type of duality can be extended to the case of arbitrary D$p$-branes ($p\neq3$), for which the 
world-volume gauge theory is again equivalent to the supergravity on the near-horizon
background generated by the D$p$-branes \cite{Itzhaki:1998dd}. The dual gauge theory is a $(p+1)$-dimensional
$U(N)$ supersymmetric Yang-Mills theory. Contrarily to the case of the D$3$-branes, it has a dimensionful 
coupling constant and the effective coupling depends on the energy scale: the gauge theory is no longer 
conformal. 

In such cases, holography has not been explored very extensively 
\cite{Sekino:1999av, Cai:1999xg, Sekino:2000mg, Gherghetta:2001iv, Asano:2003xp, Asano:2004vj}
and just recently an exhaustive extension of the holographic renormalization procedure has been
formulated \cite{Wiseman:2008qa, Kanitscheider:2008kd, Kanitscheider:2009as}. The key point in
\cite{Kanitscheider:2008kd} is the existence of a frame \cite{Duff:1994fg} in which the near-horizon 
geometry induced by the D$p$-branes is conformally $AdS_{p+2}\times S^{8-p}$ 
\cite{Boonstra:1997dy, Boonstra:1998yu, Boonstra:1998mp}. In this frame, the existence of a generalized
conformal symmetry \cite{Jevicki:1998ub} becomes manifest and the radial direction 
(transverse to the boundary) acquires the meaning of energy scale of the dual gauge theory 
\cite{Boonstra:1998mp, Skenderis:1998dq}, as in the original $AdS/CFT$-correspondence. 
Moreover, the holographic RG flow turns out to be trivial and the theory flows just because of the 
dimensionality of the coupling constant. In the case of the D$4$-branes, the theory flows to a 
$6$-dimensional fixed point at strong coupling: the world-volume theory of D$4$-branes flows to the 
world-volume theory of M$5$-branes. In \cite{Kanitscheider:2009as} an interesting observation has been
made, which drastically simplifies the direct computation of the local counterterms. The $(p+2)$-dimensional
bulk effective action, which is obtained by Kaluza-Klein reduction on the $(8-p)$-dimensional compact
manifold, can be recovered by dimensional reduction of the theory on pure (asymptotically) $AdS_{2\sigma+1}$
on a torus. The parameter $\sigma$ is related to the power of the radial direction in the dilaton and takes 
fractional values for some $p$. One can consider $\sigma$ as a generic integer and, after the 
compactification on the torus, analytically continue it to take its actual value. A simple way to then 
compute the counterterm action is to map the problem to pure $AdS_{2\sigma+1}$ theory. The counterterms are 
therefore the ones needed to renormalize these higher dimensional pure gravity theory and it is possible to go
back to the original problem by Kaluza-Klein reduction of these counterterms on a torus with a warp factor 
dependent on the dilaton.

Gauge/gravity correspondence can be further generalized by inserting extra degrees of freedom in the theory.
More precisely, one can add a finite number of branes and consider the probe approximation, so that the
backreaction on the background geometry can be neglected. Inserting probe branes introduces a fundamental 
hypermultiplet in the gauge theory, partially or completely breaking the original supersymmetries
\cite{Karch:2002sh}. Also in the case of probe-brane modes, there are IR-divergences appearing and, 
therefore, a consistent extension of the holographic renormalization procedure is needed. Such an
extension is straightforward for the case of probe-branes in a D$3$-brane background, but still shows
a very interesting feature \cite{Karch:2005ms}. The probe branes wrap an $AdS_{5-k}\times S^{3-k}$ subspace
of the whole $AdS_{5}\times S^{5}$ space-time, where $k$ is the codimension of the defect ($k\,=\,0,1,2$).
For $k\,\neq\,0$, there are two different ways to describe the embedding of the probe branes: one can fix
the position of the probe branes in the transverse space and study the embedding of the branes inside
$AdS_{5}$, where they wrap an $AdS_{5-k}$ submanifold (linear embedding)\footnote{For $k=1$ this description 
is non-supersymmetric and corresponds to turn on a vev for the embedding mode \cite{Skenderis:2002vf}.}; 
the other possible choice is to fix 
the position of the probe-branes in $AdS_{5}$ and consider the embedding of the branes in the transverse space
(angular embedding)\footnote{Choosing this embedding corresponds in turning on a massive deformation, if 
$k=0,1$. For $k=2$, it has been shown to be equivalent to turning on a vev deformation \cite{Karch:2005ms}.
This is a due to the fact that only in this case the mode saturates the Breitenlohner-Freedman 
bound \cite{Breitenlohner:1982bm}, consequently changing the boundary expansion.}. 
For $k=0$ only the latter description is possible since, in this case, the probe-branes 
wrap the whole $AdS_{5}$-space. In \cite{Karch:2005ms} it has been shown that the angular embedding mode
behaves in a neighbourhood of the boundary as a free massive scalar propagating in $AdS_{p+2-k}$, i.e.
near the boundary it has the same expansion of a free massive scalar for all the relevant orders. 
This implies that the counterterms needed to holographically renormalize these degrees of freedom
are just the same counterterms needed for a free massive scalar in $AdS$-space.

In this paper we discuss holographic renormalization of probe branes in general D$p$-brane backgrounds.
Systems of this type have been studied especially in attempt to try to 
construct holographic duals of large-$N$ QCD 
\cite{Wang:2003yc, Kruczenski:2003uq, Sakai:2004cn, Sakai:2005yt, Burrington:2009fm} and may be potentially
interesting to infer features of condensed matter systems. 
We begin with the detailed analysis of the systems with a codimension-one defect and analyze both of the
two classes of embeddings. The linear embedding is straightforward to treat. The action shows just a single
term which diverges as the boundary is approached, and it is renormalized by a term proportional to
the volume of the boundary of a warped $AdS$-space.

The angular embedding description is also very interesting. As for the conformal case \cite{Karch:2005ms},
the expansion of the mode near the boundary turns out to be the same of a free massive scalar propagating
in $AdS$. This $AdS$-space is higher dimensional with respect to the conformally-$AdS$ space that
the probe branes wrap. These ``extra''-dimensions are due to the leading contribution of the dilaton and
their number can be fractional, as in the case of the theory with no flavour \cite{Kanitscheider:2009as}.
We show that the DBI-action for the probe branes in D$p$-branes background can be equivalently rewritten as
a DBI-action in an $AdS_{q+1}$ geometry, $q$ being initially an arbitrary integer. From this perspective, the
embedding mode behaves as a free massive scalar as mentioned earlier, and the Breitenlohner-Freedman
bound is strictly satisfied for any $p$. The counterterms which renormalize the action are therefore
the ones needed by a free massive scalar particle in $AdS_{q+1}$. In order to restore the original setup,
we can compactify the $AdS_{q+1}$ on a $T^{q-p}$ torus and analytically continue $q$ to its actual
value $q_{p}$. The torus $T^{q-p}$ has a warp factor which again depends on the dilaton. However, in this
case one needs to take into account not just the dilaton factor coming from the induced metric on the 
world-volume, but also the one contained in the original DBI-action. We also apply the renormalized action 
obtained to compute the one point correlator for the operator dual to the embedding modes. 
These procedure can be straightforwardly extended to brane intersections with codimension-$0$ and 
codimension-$2$. As for the D$3$/D$3$ system \cite{Karch:2005ms}, all the D$p$/D$p$ systems ($p<5$)
show a different counterterm structure with respect to theories with a lower-codimension defect:
this is the only case for which a term proportional to $(\log{\epsilon})^{-1}$ appears. It is
a direct consequence of the fact that the embedding mode saturates the Breintelhoner-Freedman bound, which
also imply that the one-point correlator for the dual operator is expressed through the coefficient
of the normalizable mode.

This viewpoint can be extended also to the linear embedding case treated earlier. Again, one can
show the equivalence between the DBI-action of probe-branes embedded in D$p$-backgrounds and
the DBI-action of probe branes embedded into higher dimensional $AdS$-space. The only counterterm
needed is proportional to the volume of the boundary of the higher dimensional $AdS$-space and its
Kaluza-Klein reduction
on a $T^{q-p}$ returns the same counterterm found before. This case is indeed very simple by itself,
so the higher-dimensional view-point is not strictly needed. However, it contributes to 
a more general understanding of the structure of the systems we are considering. The idea
proposed in \cite{Kanitscheider:2009as} can therefore be extended to the case of non-conformal
systems with flavours.

The paper is organized as follows. In section \ref{HolRen}, we briefly review the holographic
renormalization procedure for general D$p$-backgrounds. In section \ref{Cod1} we introduce
the D$p$/D$(p+2)$ systems and we generally discuss it taking into consideration all
the possible description for the embedding of the probe D$(p+2)$-branes. In section 
\ref{LinEmb} we fix the probe branes to wrap the maximal sphere $S^{2}\subset S^{8-p}$ and consider
the embedding of the probe branes into the $(p+2)$-dimensional non-compact manifold. We show that
there is a single counterterm needed for the renormalization of the action and outline
the first suggestions about possible relations between these systems and probe branes
in a higher dimensional $AdS$-space. We also compute the one-point correlator.
In section \ref{AngEmb}, we show that,the expansion near the boundary of the angular embedding function
satisfies the equation of motion for a free massive scalar in a higher-dimensional $AdS$ space-time 
for all the orders of interest and we implement the procedure outlined earlier for the computation
of the counterterms. Furthermore, we compute the one-point correlator for the related boundary operator.
In section \ref{Codk} we extend this approach to general brane intersections. Section \ref{Conc} contains
conclusion and a summary of the results.

\section{Holographic Renormalization of D$p$-branes background}\label{HolRen}

%In this section we briefly review the holographic renormalization procedure for D$p$-brane. 
Let us start with recalling the brane solution from type IIA/IIB string theory in Euclidean signature
\begin{equation}\eqlabel{Brane}
 ds_{\mbox{\tiny $10$}}^2\:=\:\left(1+\frac{r_{p}^{7-p}}{r^{7-p}}\right)^{-1/2}\delta_{\mu\nu}dx^{\mu}dx^{\nu}+
  \left(1+\frac{r_{p}^{7-p}}{r^{7-p}}\right)^{1/2}ds^{2}_{\mbox{\tiny T}},
\end{equation}
where $\mu,\nu\,=\,0,\ldots,p$, $ds^{2}_{\mbox{\tiny T}}$ is the line element for the transverse space and
the constant $r_{p}$ is defined through
\begin{equation}\eqlabel{rp}
r_{p}^{7-p}\:\overset{\mbox{\tiny def}}{=}\left(2\sqrt{\pi}\right)^{5-p}\Gamma\left(\frac{7-p}{2}\right)g_{\mbox{\tiny s}}
 N\left(\alpha'\right)^{(7-p)/2}\:\equiv\:d_{\mbox{\tiny p}}g_{\mbox{\tiny s}}N\left(\alpha'\right)^{(7-p)/2}.
\end{equation}
The decoupling limit
\begin{equation}\eqlabel{DecLim}
 g_{\mbox{\tiny s}}\:\rightarrow\:0,\quad
 \alpha'\:\rightarrow\:0,\quad
 U\:\overset{\mbox{\tiny def}}{=}\frac{r}{\alpha'}\equiv\mbox{ fixed},\quad
 g_{\mbox{\tiny YM}}^2 N\equiv\mbox{ fixed},
\end{equation}
where the coupling constant $ g_{\mbox{\tiny YM}}$ is dimensionful and defined by
\begin{equation}\eqlabel{gYM}
 g_{\mbox{\tiny YM}}^2\:\overset{\mbox{\tiny def}}{=}\:g_{\mbox{\tiny s}}\left(2\pi\right)^{p-2}
  \left(\alpha'\right)^{(p-3)/2},
\end{equation}
corresponds to the near horizon geometry for the D$p$-branes
\begin{equation}\eqlabel{Dp}
 \begin{split}
  ds_{\mbox{\tiny $10$}}^2\:&=\:g_{\mbox{\tiny MN}}dx^{\mbox{\tiny M}}dx^{\mbox{\tiny N}}\:=\\
 &=\:\alpha'
  \left\{
   \left(\frac{U}{U_{\mbox{\tiny $p$}}}\right)^{(7-p)/2}\delta_{\mu\nu}dx^{\mu}dx^{\nu}+
   \left(\frac{U_{\mbox{\tiny $p$}}}{U}\right)^{(7-p)/2}\left[dU^2+U^2 \,d\Omega_{\mbox{\tiny $8-p$}}^2\right]\right\},
 \end{split}
\end{equation}
where the constant $U_{\mbox{\tiny $p$}}$ has been defined as
\begin{equation}\eqlabel{Up}
 U_{\mbox{\tiny $p$}}^{7-p}\:\overset{\mbox{\tiny def}}{=}\:
  \frac{d_{\mbox{\tiny $p$}}}{\left(2\pi\right)^{p-2}}g_{\mbox{\tiny YM}}^{2}N,
\end{equation}
while the dilaton and the background $(p+1)$-form are respectively
\begin{equation}\eqlabel{DilForm}
 e^{\phi}\:=\:
  \frac{g_{\mbox{\tiny YM}}^2 N}{\left(2\pi\right)^{p-2}N}\left(\frac{U}{U_{\mbox{\tiny $p$}}}\right)^{(7-p)(p-3)/4},
  \quad
  C_{\mbox{\tiny $0\ldots p$}}\:=\:\frac{\left(2\pi\right)^{p-2}\left(\alpha'\right)^{(p+1)/2}N}{g_{\mbox{\tiny YM}}^2 N}
   \left(\frac{U}{U_{\mbox{\tiny $p$}}}\right)^{7-p}.
\end{equation}
As we just mentioned, for $p\,\neq\,3$ the coupling constant is dimensionful and, therefore, the effective
coupling turns out to run with the energy scale:
\begin{equation}\eqlabel{geff}
 g_{\mbox{\tiny eff}}^{2}\:=\:g_{\mbox{\tiny YM}}^{2} N U^{p-3}.
\end{equation}
The background metric \eqref{Dp} is actually conformal to an 
$AdS_{\mbox{\tiny $p+2$}}\times S^{\mbox{\tiny $8-p$}}$ space for $p\,\neq\,5$ 
\cite{Boonstra:1998mp, Skenderis:1998dq}. 
This can be easily seen by redefining the radial coordinate according to
\begin{equation}\eqlabel{radu}
 \frac{u^2}{u_{\mbox{\tiny $p$}}^2}\:\overset{\mbox{\tiny def}}{=}\:
  \left(\frac{d_{\mbox{\tiny $p$}}}{\left(2\pi\right)^{p-2}}g_{\mbox{\tiny YM}}^2 N\right)^{-1}U^{5-p},
  \qquad
  u_{\mbox{\tiny p}}\:=\:\frac{5-p}{2}
\end{equation}
and rewriting the line element \eqref{Dp} as
\begin{equation}\eqlabel{DpConf}
  ds_{\mbox{\tiny $10$}}^{2}\:=\:\left(N\,e^{\phi}\right)^{2/(7-p)}d\hat{s}_{\mbox{\tiny $10$}}^2,
\end{equation}
so that the line element $d\hat{s}_{\mbox{\tiny $10$}}^2$ describes an $AdS_{p+2}\times S^{8-p}$ geometry
in the Poincar{\'e} patch
\begin{equation}\eqlabel{hDp}
 \begin{split}
  d\hat{s}_{\mbox{\tiny $10$}}^2\:=\:\hat{g}_{\mbox{\tiny MN}}^2 dx^{\mbox{\tiny M}}dx^{\mbox{\tiny N}}\:=
  \:\alpha'\,B_{\mbox{\tiny $p$}}\left\{u^2\,\delta_{\mu\nu}dx^{\mu}dx^{\nu}+\frac{du^2}{u^2}+
   u_{p}^2\,d\Omega_{\mbox{\tiny $8-p$}}^2\right\}.
 \end{split}
\end{equation}
The near-horizon geometry of a background generated by a stack of $N$ D$p$-branes is therefore conformal to
a $AdS_{p+2}\times S^{8-p}$ space with conformal factor $\left(N e^{\phi}\right)^{\frac{2}{7-p}}$.
In the coordinates $\left\{x^{\mu},\,u\right\}$, the dilaton writes
\begin{equation}\eqlabel{Dilu}
 e^{\phi}\:=\:\frac{A_{\mbox{\tiny $p$}}}{N}\,u^{(p-3)(7-p)/2(5-p)}.
\end{equation}
For the purpose of holographic renormalization, it is convenient to rewrite the metric \eqref{hDp} in 
Fefferman-Graham coordinates through the radial coordinate redefinition $\rho\,=\,u^{-1}$. The original 
ten-dimensional metric \eqref{Dp} acquires the form
\begin{equation}\eqlabel{Dp2}
 \begin{split}
  ds_{\mbox{\tiny $10$}}^2\:&=\:\alpha'\,B_{\mbox{\tiny $p$}}\left(N\,e^{\phi}\right)^{2/(7-p)}
   \left\{
    \frac{\delta_{\mu\nu}dx^{\mu}dx^{\nu}+d\rho^2}{\rho^2}+u_{p}^2\,d\Omega_{\mbox{\tiny $8-p$}}^2
   \right\}\:\equiv\\
   &\equiv\:\alpha'\,B_{\mbox{\tiny $p$}}\left(N\,e^{\phi}\right)^{2/(7-p)}d\tilde{s}^{2}_{\mbox{\tiny $10$}},
 \end{split}
\end{equation}
with 
\begin{equation}\eqlabel{Dilr}
e^{\phi}\:=\:\frac{A_{\mbox{\tiny $p$}}}{N}\rho^{-(p-3)(7-p)/2(5-p)}.
\end{equation}
For the sake of generality, let us rewrite the line element $d\tilde{s}_{\mbox{\tiny $10$}}^2$ in the 
following form
\begin{equation}\eqlabel{FGmetric10}
 d\tilde{s}_{\mbox{\tiny $10$}}^2\:=\:
   \mathfrak{g}_{\mbox{\tiny $\hat{\mu}\hat{\nu}$}}dx^{\tilde{\mu}}dx^{\tilde{\nu}}+
    u_{p}^2 d\Omega^{2}_{8-p}\:=\:
   \frac{\mathtt{g}_{\mu\nu}\left(x,\rho\right)dx^{\mu}dx^{\nu}+d\rho^2}{\rho^2}+
    u_{p}^2 d\Omega^{2}_{8-p}
\end{equation}
In \cite{Kanitscheider:2008kd} it was shown that in a neighbourhood of the boundary the expansion for
both the metric $\mathtt{g}_{\mu\nu}(x,\rho)$ and the dilaton $\phi(x,\rho)$ may contain fractional powers
of $\rho$:
\begin{equation}\eqlabel{GDexp}
 \begin{split}
  &\mathtt{g}_{\mu\nu}\left(x,\rho\right)\:=\:
    \mathtt{g}^{\mbox{\tiny $\left(0\right)$}}_{\mu\nu}\,+\,
    \rho^{2}\mathtt{g}^{\mbox{\tiny $\left(2\right)$}}_{\mu\nu}\,+\,\ldots\,+\,
    \rho^{2\frac{7-p}{5-p}}
    \left[
     \mathtt{g}^{\mbox{\tiny $\left(2\frac{7-p}{5-p}\right)$}}_{\mu\nu}+
     \left(\delta_{\mbox{\tiny $p,3$}}+\delta_{\mbox{\tiny $p,4$}}\right)
      \mathtt{h}^{\mbox{\tiny $\left(2\frac{7-p}{5-p}\right)$}}_{\mu\nu}\log{\rho}
    \right]\,+\,\ldots\\
  &\phi\left(x,\rho\right)\:=\:
    -\frac{\left(p-3\right)\left(7-p\right)}{2\left(5-p\right)}\log{\rho}\,+\,
    \left(\varepsilon_{\mbox{\tiny $p,3$}}\frac{7-p}{p-3}+\delta_{\mbox{\tiny $p,3$}}\right)\times\\
  &\hspace{1.25cm}
   \times
    \left\{
     \kappa^{\mbox{\tiny $\left(0\right)$}}\,+\,
     \rho^{2}\kappa^{\mbox{\tiny $\left(2\right)$}}\,+\,\ldots\,+\,
     \rho^{2\frac{7-p}{5-p}}
     \left[
      \kappa^{\mbox{\tiny $\left(2\frac{7-p}{5-p}\right)$}}+
      \left(\delta_{\mbox{\tiny $p,3$}}+\delta_{\mbox{\tiny $p,4$}}\right)
       \mathtt{k}^{\mbox{\tiny $\left(2\frac{7-p}{5-p}\right)$}}\log{\rho}
     \right]\,+\,\ldots
    \right\},
 \end{split}
\end{equation}
where $\varepsilon_{\mbox{\tiny $p,3$}}$ is $0$ for $p=3$ and $1$ otherwise. The undetermined
coefficients of these expansions appear at order $\mathcal{O}\left(\rho^0\right)$ and
$\mathcal{O}\left(\rho^{2\frac{7-p}{5-p}}\right)$. Furthermore, the boundary
expansions \eqref{GDexp} exhibit a behaviour similar to asymptotically $AdS$ backgrounds. More
specifically, the D$p$ backgrounds with $p<3$ do not show any logarithmic term, as for 
asymptotically $AdS$-spaces with even dimensions. For $p=3,4$, which are the only values of $p$ for which
the last term in \eqref{GDexp} appears at an even integer power of $\rho$, they behave as for
asymptotically $AdS$-spaces with odd dimensions. In \cite{Kanitscheider:2009as}, it was explicitly shown
the equivalence between the effective supergravity action which allows for D$p$-brane solutions 
(compactified on $S^{8-p}$) and the action for a theory on pure $AdS_{2\sigma+1}$ with cosmological constant 
$\Lambda\,=\,-\sigma\left(2\sigma-1\right)$ when it is compactified on a $T^{2\sigma-d}$
\begin{equation}\eqlabel{redants}
 \begin{split}
 &ds_{\mbox{\tiny $2\sigma+1$}}^2\:=\:
   \mathfrak{g}_{\mbox{\tiny $\hat{\alpha}\hat{\beta}$}}^{\mbox{\tiny $(p+2)$}}
    dx^{\hat{\alpha}}dx^{\hat{\beta}}+e^{2\frac{\phi(x,\rho)}{2\sigma-(p+1)}}\delta_{ab}dx^{a}dx^{b}\\
 &S\:=\:L_{\mbox{\tiny $AdS$}}\int d^{\mbox{\tiny $2\sigma+1$}}x\:
   \sqrt{g_{\mbox{\tiny $(2\sigma+1)$}}}
   \left[
    R^{\mbox{\tiny $(2\sigma+1)$}}+2\sigma\left(2\sigma-1\right)
   \right]\:=\\
 &\hspace{.3cm}
   =\:L
    %L_{\mbox{\tiny $AdS$}}\left(2\pi R_{\mbox{\tiny T}}\right)^{\mbox{\tiny $2\sigma-(p+1)$}}
    \int d^{p+2}x\:e^{\phi}\sqrt{\mathfrak{g}_{\mbox{\tiny $(p+2)$}}}
    \left[
     R^{\mbox{\tiny $(p+2)$}}+
     \frac{2\sigma-(p+2)}{2\sigma-(p+1)}\left(\partial\phi\right)^2+2\sigma\left(2\sigma-1\right)
    \right],
 \end{split}
\end{equation}
with the identification 
$L=L_{\mbox{\tiny $AdS$}}\left(2\pi R_{\mbox{\tiny T}}\right)^{\mbox{\tiny $2\sigma-(p+1)$}}$.
From a $(2\sigma+1)$-dimensional perspective, the action \eqref{redants} is renormalized by the standard 
$AdS$-counterterms. Considering $2\sigma$ as an arbitrary integer, one can perform the 
Kaluza-Klein reduction of the $AdS$-counterterms with the metric ansatz \eqref{redants} and
then analytically continue $2\sigma$ to take the fractional value 
$2\sigma_{\mbox{\tiny $p$}}=2(7-p)/(5-p)$. The counterterms for pure $AdS$-gravity are 
\cite{deHaro:2000xn}
\begin{equation}\eqlabel{AdSct}
 \begin{split}
  S_{\mbox{\tiny $ct$}}\:=\:&
   L_{\mbox{\tiny $AdS$}}\int_{\rho=\epsilon}d^{2\sigma}x\sqrt{\gamma_{\mbox{\tiny $(2\sigma)$}}}
    \left[
     2(2\sigma-1)+\frac{1}{2\sigma-2}R_{\mbox{\tiny $(\gamma)$}}+\right.\\
     &+\left.
     \frac{1}{(2\sigma-4)(2\sigma-2)^2}
     \left(
      R_{\mbox{\tiny $AB$}}^{\mbox{\tiny $(\gamma)$}}
       R^{\mbox{\tiny $AB$}}_{\mbox{\tiny $(\gamma)$}}-
      \frac{\sigma}{2(2\sigma-1)}R_{\mbox{\tiny $(\gamma)$}}^2
     \right)-
     a_{\mbox{\tiny $(2\sigma)$}}\log{\epsilon}+\ldots
    \right],
 \end{split}
\end{equation}
with $a_{\mbox{\tiny $(2\sigma)$}}$ indicating the conformal anomaly. Dimensional reducing \eqref{AdSct}
on the metric ansatz \eqref{redants}, one obtain the counterterms for the $(p+2)$-dimensional background
\cite{Kanitscheider:2008kd, Kanitscheider:2009as}
\begin{equation}\eqlabel{DpCt}
 \begin{split}
   S_{\mbox{\tiny $ct$}}\:=\:&L\int_{\rho=\epsilon}d^{p+2}x\,e^{\phi}\sqrt{\gamma_{\mbox{\tiny $(p+1)$}}}
    \left\{
     2(2\sigma-1)+
     \frac{1}{2\sigma-2}
     \left(
      R_{\mbox{\tiny $(p+1)$}}+\frac{2\sigma-p-2}{2\sigma-p-1}\left(\partial\phi\right)^2
     \right)+
    \right.\\
   &+
    \delta_{\mbox{\tiny $\sigma,3$}}
    \left[
     \frac{1}{(2\sigma-4)(2\sigma-2)^2}
     \left(
      R_{\mbox{\tiny $\alpha\beta$}}^{\mbox{\tiny $(p+1)$}}-
      2\frac{p-3}{7-p}
      \left(
       \nabla_{\mbox{\tiny $\alpha$}}\partial_{\mbox{\tiny $\beta$}}\phi+
       \partial_{\mbox{\tiny $\alpha$}}\phi\partial_{\mbox{\tiny $\beta$}}\phi
      \right)
     \right)^2+
    \right.\\
   &\hspace{1.5cm}+
     \frac{1}{(2\sigma-4)(\sigma-1)^2}\frac{p-3}{7-p}
     \left(\Box_{\mbox{\tiny $(p+1)$}}\phi+\left(\partial\phi\right)\right)^2-\\
   &-\left.\left.
     \frac{\sigma}{(2\sigma-4)(2\sigma-2)^2}
     \left(
      R_{\mbox{\tiny $(p+1)$}}-4\frac{p-3}{7-p}
      \left(\Box_{\mbox{\tiny $(p+1)$}}\phi+\left(\partial\phi\right)^2\right)
     \right)^2+
     a_{\mbox{\tiny $(2\sigma)$}}\log{\epsilon}
    \right]
   \right\}.
 \end{split}
\end{equation}
The explicit expression for the conformal anomaly $a_{\mbox{\tiny $(2\sigma)$}}$ is provided in 
\cite{Kanitscheider:2008kd}.

The renormalized action is then given by
\begin{equation}\eqlabel{Sren}
 S_{\mbox{\tiny ren}}\:=\:\lim_{\epsilon\rightarrow0}
  \left[S|_{\mbox{\tiny $\epsilon$}}+S_{\mbox{\tiny GH}}+S_{\mbox{\tiny ct}}\right],
\end{equation}
where $S_{\mbox{\tiny GH}}$ is the standard Gibbons-Hawking term which needs to be introduced in
order to have a well-defined variational principle. From an $AdS$ (and therefore conformal) point of
view, the one-point correlator for the boundary stress-energy tensor is 
\cite{Balasubramanian:1999re, deHaro:2000xn}
\begin{equation}\eqlabel{1ptT}
 \langle T_{\mbox{\tiny $AB$}}\rangle_{\mbox{\tiny $(2\sigma)$}}\:=\:
  \frac{2}{\sqrt{\gamma_{\mbox{\tiny $(2\sigma)$}}}}
  \frac{\delta S_{\mbox{\tiny ren}}}{\delta\gamma^{\mbox{\tiny $\alpha\beta$}}_{\mbox{\tiny $(2\sigma)$}}}.
\end{equation}
The dimensional reduction of \eqref{1ptT} on the torus $T^{\mbox{\tiny $2\sigma-p-1$}}$ returns both the
one-point correlator for $(p+1)$-dimensional boundary stress-energy tensor and the one for
the scalar operator dual to the dilaton field:
\begin{equation}\eqlabel{1ptT2}
 \begin{split}
 &e^{\kappa_{\mbox{\tiny $(0)$}}}\left(2\pi R_{\mbox{\tiny T}}\right)^{2\sigma-p-1}
 \langle T_{\mbox{\tiny $\alpha\beta$}}\rangle_{\mbox{\tiny $(2\sigma)$}}\:=\:
 2\sigma L e^{\kappa_{\mbox{\tiny $(0)$}}} g_{\mbox{\tiny $\alpha\beta$}}^{\mbox{\tiny $(2\sigma)$}}
 \:=\:\langle T_{\alpha\beta}\rangle_{\mbox{\tiny $(p+1)$}}\\
 &e^{\kappa_{\mbox{\tiny $(0)$}}}\left(2\pi R_{\mbox{\tiny T}}\right)^{2\sigma-p-1}
 \langle T_{ab}\rangle_{\mbox{\tiny $(2\sigma)$}}\:=\:
 \frac{4\sigma L}{2\sigma-p-1}e^{\frac{2\sigma-p+1}{2\sigma-p-1}\kappa_{\mbox{\tiny $(0)$}}}
 \kappa_{\mbox{\tiny $(2\sigma)$}}\delta_{ab}\:=\:
 -e^{\frac{2}{2\sigma-p-1}\phi}\delta_{ab}\langle\mathcal{O}_{\phi}\rangle_{\mbox{\tiny $(p+1)$}}
 \end{split}
\end{equation}

The remarkable observation of \cite{Kanitscheider:2009as} that non-conformal backgrounds can 
be mapped into higher dimensional asymptotically $AdS$-geometries drastically simplifies the study of the 
dynamics of such systems, which may be determined in terms of the dynamics of conformal systems.

\section{Brane intersections with codimension-1 defect}\label{Cod1}

In the background \eqref{Dp2} we introduce $M$ parallel probe D$(p+2)$-branes ($M\ll N$) according to the 
following intersection configuration

\begin{center}
\begin{tabular}{||p{2cm}||*{15}{c|}|}
 \hline
           &  0  &  1  &  2  &  \ldots  &  p-2  & p-1 &  p  & p+1 & p+2 & p+3 & p+4 & \ldots &  8  &  9  \\
 \hline
 \hline
 D$p$      &  X  &  X  &  X  &  \ldots  &   X   &  X  &  X  & { } & { } & { } & { } & \ldots & { } & { } \\
 \hline
 D$(p+2)$  &  X  &  X  &  X  &  \ldots  &   X   &  X  & { } &  X  &  X  &  X  & { } & \ldots & { } & { } \\
 \hline
\end{tabular}
\end{center}

The probe branes wrap an internal 2-sphere $S^{2}\subset S^{8-p}$. Given the presence of a codimension-1 
defect, the embedding of the D$(p+2)$ branes in the D$p$-brane background can in principle be described 
through two functions $x^{p}\,\equiv\,z(\rho)$ and $\theta\,\equiv\,\theta(\rho)$, where $\theta$ is one of 
the angular coordinates of $S^{8-p}$:
\begin{equation}\eqlabel{S8-p}
 d\Omega_{\mbox{\tiny $8-p$}}^2\:=\:
  d\theta^2 + \sin^2{\theta}d\Omega_{2}^2 + \cos^{2}{\theta}d\Omega_{\mbox{\tiny $5-p$}}^2
\end{equation}
For the moment, let us keep both of the two embedding functions and consider the pure geometrical case, 
in which the probe branes do not carry any gauge field. The action for the probe branes is given only by the 
DBI-term
\begin{equation}\eqlabel{DBI}
 \begin{split}
  S_{\mbox{\tiny D$(p+2)$}}\:&=\:M\,T_{\mbox{\tiny D(p+2)}}\int\:d^{p+3}\xi\:
    e^{-\phi}\sqrt{g_{\mbox{\tiny $p+3$}}}\:=\\
   &=\:M\,T_{\mbox{\tiny D(p+2)}}\mathcal{N}_{\mbox{\tiny $p$}}
      \int d^{p+3}\xi\:e^{2\frac{p-2}{7-p}\phi}\sqrt{\tilde{g}_{\mbox{\tiny $p+3$}}}
 \end{split}
\end{equation}
where
\begin{equation}\eqlabel{N}
\mathcal{N}_{\mbox{\tiny $p$}}\:\overset{\mbox{\tiny def}}{=}\:
 \left(\alpha'\right)^{(p+3)/2}B_{\mbox{\tiny $p$}}^{(p+3)/2}N^{(p+3)/(7-p)}
\end{equation}
and $g_{\mbox{\tiny $p+3$}}$ and $\tilde{g}_{\mbox{\tiny $p+3$}}$ are the determinants of the world-volume 
metric induced by the background metrics $g_{\mbox{\tiny MN}}$ and $\tilde{g}_{\mbox{\tiny MN}}$ respectively. The price one pays in changing
frame, beside the overall constant $\mathcal{N}_{\mbox{\tiny p}}$, is a shift in the dilaton factor of the world-volume action
\begin{equation}\eqlabel{DilFac}
 e^{-\phi}\:\longrightarrow\:e^{2\frac{p-2}{7-p}\phi},
\end{equation}
which substantially leaves the structure of the action unchanged. It is interesting to notice that, in the 
frame we are now considering, there is no non-trivial dilaton dependence in \eqref{DBI} for $p\,=\,2,\,3$.
The induced metric on the D$(p+2)$-brane world-volume is
\begin{equation}\eqlabel{IndM}
 d\tilde{s}_{\mbox{\tiny $p+3$}}^2\:=\:\frac{\delta_{\alpha\beta}}{\rho^2}dx^{\alpha}dx^{\beta}+
  \left[1+\left(z'\right)^2+u_{\mbox{\tiny $p$}}^2 \rho^2\left(\theta'\right)^2\right]\frac{d\rho^2}{\rho^2}+
  +u_{\mbox{\tiny $p$}}^2 \sin^{2}{\theta}d\Omega_{2}^2,
\end{equation}
where the indices $\alpha,\,\beta\,=\,0,\ldots,p-1$ and the prime $'$ indicates the first derivative with 
respect to the radial coordinate $\rho$. The action \eqref{DBI} can be easily integrate over the $S^{2}$ 
coordinates to give
\begin{equation}\eqlabel{DBI2}
 \begin{split}
  S_{\mbox{\tiny D$(p+2)$}}\:
   &=\:M\,T_{\mbox{\tiny D(p+2)}}\,\hat{\mathcal{N}}_{\mbox{\tiny $p$}}
        \int dt\,d^{p-1}x\,d\rho\,e^{2\frac{p-2}{7-p}\phi}\,\sin^{2}{\theta}\sqrt{\tilde{g}_{\mbox{\tiny $p+1$}}}\\
   &=\:M\,T_{\mbox{\tiny D(p+2)}}\,\hat{\mathcal{N}}_{\mbox{\tiny $p$}}
        \int dt\,d^{p-1}x\,\frac{d\rho}{\rho^{p+1}}\,e^{2\frac{p-2}{7-p}\phi}\,\sin^{2}{\theta}
	 \sqrt{1+\rho^{-2}\left(\partial z\right)^{2}+
	  %\tilde{g}^{\mbox{\tiny $\hat{\alpha}\hat{\beta}$}}
	  %\left(\partial_{\mbox{\tiny $\hat{\alpha}$}}z\right)
	  %\left(\partial_{\mbox{\tiny $\hat{\beta}$}}z\right)+
	  u_{\mbox{\tiny $p$}}^2 \left(\partial\theta\right)^{2}
	  %\tilde{g}^{\mbox{\tiny $\hat{\alpha}\hat{\beta}$}}
	  %\left(\partial_{\mbox{\tiny $\hat{\alpha}$}}\theta\right)
	  %\left(\partial_{\mbox{\tiny $\hat{\beta}$}}\theta\right)}
	  },
 \end{split}
\end{equation}
with $x^{\mbox{\tiny $\hat{\alpha}$}}\,=\,\left\{x^{\mbox{\tiny $\alpha$}},\rho\right\}$,
$\hat{\mathcal{N}}_{\mbox{\tiny $p$}}\,=\,
 u_{\mbox{\tiny $p$}}^2\,\mbox{vol}\left\{S^{2}\right\}\mathcal{N}_{\mbox{\tiny $p$}}$,
$\left(\partial f\right)^2\,\equiv\, 
     \mathfrak{g}^{\mbox{\tiny $\hat{\alpha}\hat{\beta}$}}
     \left(\partial_{\mbox{\tiny $\hat{\alpha}$}}f\right)
     \left(\partial_{\mbox{\tiny $\hat{\beta}$}}f\right)$ and
\begin{equation}\eqlabel{FGmetric}
 \mathfrak{g}_{\mbox{\tiny $\tilde{\alpha}\tilde{\beta}$}}dx^{\tilde{\alpha}}dx^{\tilde{\beta}}
 \:\overset{\mbox{\tiny def}}{=}\:\frac{\delta_{\alpha\beta}dx^{\alpha}dx^{\beta}+d\rho^2}{\rho^2}.
\end{equation}
The action \eqref{DBI2} depends on the linear embedding function $z(\rho)$ through its first derivative only. This implies that there is a first integral of motion $c_{z}$ related to it
\begin{equation}\eqlabel{zIntMot}
 c_{z}\:=\:\frac{e^{2\frac{p-2}{7-p}\phi}}{{\rho^{p+1}}}\sin^{2}{\theta}
   \frac{z'}{\sqrt{1+\rho^{-2}\left(\partial z\right)^2 + u_{\mbox{\tiny $p$}}^2\left(\partial\theta\right)^2}}.
\end{equation}
The equation of motion for both the embedding functions $z(\rho)$ and $\theta(\rho)$ are
\begin{equation}\eqlabel{EOM}
 \begin{split}
  &z'(\rho)\:=\:c_{z}\frac{\rho^{\frac{11-p}{5-p}}}{\sqrt{\sin^{2}{\theta}-c_{z}^{2}\rho^{2\frac{11-p}{5-p}}}}
                \sqrt{1+ u_{\mbox{\tiny $p$}}^2\left(\partial\theta\right)^2}\\
  &0\:=\:\Box\theta-
         \cot{\theta}\frac{c_{z}^2 \rho^{2\frac{11-p}{5-p}}}{\sin^2{\theta}-c_{z}^{2}\rho^{2\frac{11-p}{5-p}}}\,
          \left(\partial\theta\right)^2-
	 \frac{1}{2}\mathfrak{g}^{\mbox{\tiny $\hat{\alpha}\hat{\beta}$}}
	  \left(\partial_{\mbox{\tiny $\hat{\alpha}$}}\theta\right)
	  \frac{\partial_{\mbox{\tiny $\hat{\beta}$}}\left[1+
           u_{\mbox{\tiny $p$}}^2\left(\partial\theta\right)^2\right]}{1+
	    u_{\mbox{\tiny $p$}}^2\left(\partial\theta\right)^2}+\\
  &\hspace{1cm}+\left[\frac{11-p}{5-p}\frac{c_{z}^2 \rho^{\frac{17-p}{5-p}}}{\sin^{2}{\theta}-
           c_{z}^2 \rho^{2\frac{11-p}{5-p}}}+\frac{(p-2)(p-3)}{5-p}\right]
	   %\delta^{\rho}_{\phantom{\rho}\mbox{\tiny $\hat{\alpha}$}}
           \mathfrak{g}^{\mbox{\tiny $\hat{\alpha}\hat{\beta}$}}
           \left(\partial_{\mbox{\tiny $\hat{\alpha}$}}\phi\right)
	   \left(\partial_{\mbox{\tiny $\hat{\beta}$}}\theta\right)
	   -\frac{2}{u_{\mbox{\tiny $p$}}^2}\cot{\theta}\frac{\sin^{2}{\theta}}{\sin^{2}{\theta}-c_{z}^{2}\rho^{2\frac{11-p}{5-p}}}
 \end{split}
\end{equation}
In \eqref{EOM}, the operator $\Box$ is constructed through the metric $\mathfrak{g}_{\mbox{\tiny $\hat{\alpha}\hat{\beta}$}}$. The case $c_{z}\,=\,0$ corresponds to the case in which the probe branes bend in the 
transverse space only.
Notice that at the boundary, the angular coordinate $\theta$ takes the value $\pi/2$, as one can 
straightforwardly see from the \eqref{EOM}. In what follows, we discuss the two different classes of 
probe-branes embeddings separately.

\section{Linear Embedding}\label{LinEmb}

The simplest case is provided by the description of the embedding of the probe branes through the linear 
coordinate $z(\rho)$. The probe branes wrap the maximal sphere $S^{2}\subset S^{8-p}$ located at 
$\theta=\pi/2$, and the scalar $z(\rho)$ describes the embedding of the branes in the $(p+2)$-dimensional 
(conformally)-$AdS$ manifold, where they wrap a $(p+1)$-dimensional (conformally)-$AdS$ subspace. 
The action and the equation of motion for the scalar $z(\rho)$ can be obtained from \eqref{DBI2} and 
\eqref{EOM} by setting $\theta\,=\,\pi/2$
\begin{equation}\eqlabel{zActEom}
 \begin{split}
  &S_{\mbox{\tiny $D(p+2)$}}^{\mbox{\tiny $\left(z\right)$}}\:=\:
        M\,T_{\mbox{\tiny D(p+2)}}\,\hat{\mathcal{N}}_{\mbox{\tiny $p$}}
        \int dt\,d^{p-1}x\,\frac{d\rho}{\rho^{p+1}}\,e^{2\frac{p-2}{7-p}\phi}\,
	 \sqrt{1+\rho^{-2}\left(\partial z\right)^{2}},\\
  &z'(\rho)\:=\:c_{z}\frac{\rho^{\frac{11-p}{5-p}}}{\sqrt{1-c_{z}^{2}\rho^{2\frac{11-p}{5-p}}}}.
 \end{split}
\end{equation}
As mentioned before, for $c_{z}\,=\,0$ the embedding function $z$ is constant and, therefore, the probe branes do not
bend. For $c_{z}\,\neq\,0$, the solution extends up to a maximum value for the radial coordinate $\rho$
\begin{equation}\eqlabel{rhomax}
 \rho_{\mbox{\tiny max}}\:=\:c^{-\frac{5-p}{11-p}}.
\end{equation}
As pointed out in \cite{Karch:2005ms} for the analysis of the D$3$/D$5$ system\footnote{The D$3$/D$5$ system belongs
to the class of theories we are considering, so the results discussed in section 5 of \cite{Karch:2005ms} are reproduced
by setting $p\,=\,3$.}, the string turns back once reaches $\rho\,=\,\rho_{\mbox{\tiny max}}$. This can be seen by expanding
the solution \eqref{zActEom} in a neighbourhood of $\rho_{\mbox{\tiny max}}$
\begin{equation}\eqlabel{rhomaxN}
 \begin{split}
  &z'\left(\rho\right)\:=\:\sqrt{\frac{5-p}{2(11-p)}\rho_{\mbox{\tiny max}}}\left(\rho_{\mbox{\tiny max}}-\rho\right)^{-1/2}
   +\ldots\\
  &z\left(\rho\right)\:=\:m+\sqrt{2\frac{5-p}{11-p}\rho_{\mbox{\tiny max}}}\left(\rho_{\mbox{\tiny max}}-\rho\right)^{1/2}
   +\ldots
 \end{split}
\end{equation}
Thus, a D$p$/D$(p+2)$ system in the probe approximation and with the embedding of the probe D$(p+2)$-branes parametrized by 
the linear coordinate $x^{p}\,\equiv\,z(\rho)$ is actually dual to defect theories of D$(p+2)$/$\bar{\mbox{D}}(p+2)$ 
separated by a finite distance proportional to $\rho_{\mbox{\tiny max}}$.

We can now focus on divergences near the boundary. Near the boundary $\rho\,=\,0$, the solution \eqref{zActEom} has the
following asymptotic expansion
\begin{equation}\eqlabel{bd}
 \begin{split}
  &z'(\rho)\:=\:c_{z}\rho^{\frac{11-p}{5-p}}\left[1+c_{z}^2 \rho^{2\frac{11-p}{5-p}}+
   \mathcal{O}\left(\rho^{4\frac{11-p}{5-p}}\right)\right]\\
  &z(\rho)\:=\:m_{\mbox{\tiny $0$}}+\frac{5-p}{2(8-p)}\rho^{2\frac{8-p}{5-p}}+
   \mathcal{O}\left(\rho^{2\frac{19-2p}{5-p}}\right)
 \end{split}
\end{equation}
From the on-shell action
\begin{equation}\eqlabel{DBIonshell}
 S_{\mbox{\tiny $D(p+2)$}}^{\mbox{\tiny $\left(z\right)$}}\Big|_{\mbox{\tiny on-shell}}\:=\:
  M\,T_{\mbox{\tiny D(p+2)}}\,\hat{\mathcal{N}}_{\mbox{\tiny $p$}}
        \int dt\,d^{p-1}x\,\int_{\epsilon}^{\rho_{\mbox{\tiny max}}}d\rho\:
	 \frac{\rho_{\mbox{\tiny max}}^{\frac{11-p}{5-p}}}{\rho^{\frac{11-p}{5-p}}}\,
	 \frac{1}{\sqrt{\rho_{\mbox{\tiny max}}^{2\frac{11-p}{5-p}}-\rho^{2\frac{11-p}{5-p}}}},
\end{equation}
where the cut-off $\epsilon$ has been introduced to regularize the action near the boundary, it's easy to 
read off the divergent boundary action
\begin{equation}\eqlabel{DBIdiv}
  S_{\mbox{\tiny $D(p+2)$}}^{\mbox{\tiny $\left(z\right)$}}\Big|_{\mbox{\tiny div}}\:=\:
   M\,T_{\mbox{\tiny D(p+2)}}\,\hat{\mathcal{N}}_{\mbox{\tiny $p$}}
    \int dt\,d^{p-1}x\:\frac{5-p}{6}\,\epsilon^{-\frac{6}{5-p}}.
\end{equation}
The existence of a single divergent term implies the need of a single counterterm to renormalize the action. 
Such a counterterm is provided by a term proportional to the volume of the boundary 
(regularized with $\rho\,=\,\epsilon$)
\begin{equation}\eqlabel{DBIct}
 S_{\mbox{\tiny $D(p+2)$}}^{\mbox{\tiny $\left(z\right)$}}\Big|_{\mbox{\tiny ct}}\:=\:
  M\,T_{\mbox{\tiny D(p+2)}}\,\hat{\mathcal{N}}_{\mbox{\tiny $p$}}
   \int dt\,d^{p-1}x\:\mathfrak{c}\,e^{2\frac{p-2}{7-p}\phi}
     \sqrt{\gamma_{\mbox{\tiny p}}\big|_{\mbox{\tiny $\epsilon$}}},
\end{equation}
with the coefficient $\mathfrak{c}$ which can be easily computed to be
\begin{equation}\eqlabel{c}
 \mathfrak{c}\:=\:-\frac{5-p}{6}.
\end{equation}
The finite on-shell action is given by the primitive function of \eqref{DBIonshell} 
(which can be computed exactly) evaluated at $\rho=\rho_{\mbox{\tiny max}}$
\begin{equation}\eqlabel{DBIfinite}
  S_{\mbox{\tiny $D(p+2)$}}^{\mbox{\tiny $\left(z\right)$}}\Big|_{\mbox{\tiny finite}}\:=\:
  M\,T_{\mbox{\tiny D(p+2)}}\,\hat{\mathcal{N}}_{\mbox{\tiny $p$}}
   \int dt\,d^{p-1}x\:\frac{p-5}{6}\rho^{-\frac{6}{5-p}}_{\mbox{\tiny max}}\varpi,
    %\left[
    % 1-\frac{3}{\sqrt{2\pi}}\sum_{k=0}^{\infty}\frac{1}{(11-p)k+(8-p)}\frac{\Gamma\left(\frac{3}{2}+k\right)}{
    %  \Gamma\left(2+k\right)}
    %\right]
\end{equation}
where $\varpi$ is a finite constant. %defined as
%\begin{equation}\eqlabel{varpi}
% \varpi\:=\: 1-\frac{3}{2\sqrt{\pi}}\sum_{k=0}^{\infty}\frac{1}{(11-p)k+(8-p)}\frac{\Gamma\left(\frac{3}{2}+k\right)}{
%              \Gamma\left(2+k\right)}.
%\end{equation}
The holographic renormalization method for the action of a probe $D(p+2)$-brane in a background generated by 
a stack of D$p$-branes thus prescribe a single counterterm \eqref{DBIct}. Thus, the renormalized action is given by
\begin{equation}\eqlabel{DBIren}
 S_{\mbox{\tiny $D(p+2)$}}^{\mbox{\tiny $\left(z\right)$}}\Big|_{\mbox{\tiny ren}}\:=\:
  \lim_{\epsilon\rightarrow0}S_{\mbox{\tiny tot}}^{\mbox{\tiny $\left(z\right)$}}(\epsilon)\:=\:
  \lim_{\epsilon\rightarrow0}
  \left[
   S_{\mbox{\tiny $D(p+2)$}}^{\mbox{\tiny $\left(z\right)$}}\Big|_{\mbox{\tiny on-shell}}+
   S_{\mbox{\tiny $D(p+2)$}}^{\mbox{\tiny $\left(z\right)$}}\Big|_{\mbox{\tiny ct}}
  \right]
\end{equation}
We can now compute the one-point correlator of the boundary operator $\mathcal{O}_{\mbox{\tiny $z$}}$ 
associated with the scalar function $z$
\begin{equation}\eqlabel{1ptcorrO}
 \langle\mathcal{O}_{\mbox{\tiny $z$}}\rangle\:=\:
  %\lim_{\epsilon\rightarrow0}\frac{1}{\epsilon^{\Delta_{\mbox{\tiny $p$}}}}
  %\frac{1}{\sqrt{\gamma|_{\mbox{\tiny $\epsilon$}}}}
  %\frac{\delta S_{\mbox{\tiny tot}}^{\mbox{\tiny $\left(z\right)$}}}{\delta z}\:=\:
  -\lim_{\epsilon\rightarrow0}
   \frac{1}{\epsilon^{\frac{11-p}{5-p}}}\left.\frac{z'}{\sqrt{1+\left(z'\right)^2}}\right|_{\mbox{\tiny $\epsilon$}}\:=\:
  -c_{\mbox{\tiny $z$}}.
\end{equation}
One comment is in order. It is remarkable that the divergent action \eqref{DBIdiv} is the same of the one
that one would obtain from probe D-branes in an $AdS$-space of dimensions $6/(5-p)+1$ and the counterterm
\eqref{DBIct} is actually given in terms of the volume of the boundary of this $AdS$-space. We will make
this observation more precise in section \ref{Conc}.

\section{Angular Embedding}\label{AngEmb}
Let us now fix the position of the probe D$(p+2)$-branes in the $(p+2)$-dimensional (conformal)-$AdS$ space 
($z=0$) and let us consider their embedding in the transverse space, which is parametrized by the angular 
coordinate $\theta(\rho)$. The action and the equation of motion can be obtain from \eqref{DBI} and \eqref{EOM} respectively by setting $c_{z}=0=z'$:
\begin{equation}\eqlabel{aActEom}
 \begin{split} 
  &S_{\mbox{\tiny D$(p+2)$}}^{\mbox{\tiny $\theta$}}\:=\:
    M\,T_{\mbox{\tiny D(p+2)}}\,\hat{\mathcal{N}}_{\mbox{\tiny $p$}}
         \int dt\,d^{p-1}x\,d\rho\,
          e^{\frac{\left(p-2\right)\left(p-3\right)}{5-p}\phi}\,\sin^{2}{\theta}\,
          \sqrt{\mathfrak{g}}\,
 	  \sqrt{1+u_{\mbox{\tiny $p$}}^2 \left(\partial\theta\right)^{2}}\\
  &0\:=\:\Box\theta-
       	  \frac{1}{2}\mathfrak{g}^{\mbox{\tiny $\hat{\alpha}\hat{\beta}$}}
	  \left(\partial_{\mbox{\tiny $\hat{\alpha}$}}\theta\right)
	  \frac{\partial_{\mbox{\tiny $\hat{\beta}$}}\left[1+
           u_{\mbox{\tiny $p$}}^2\left(\partial\theta\right)^2\right]}{1+
	    u_{\mbox{\tiny $p$}}^2\left(\partial\theta\right)^2}+
           \frac{(p-2)(p-3)}{5-p}
	   %\delta^{\rho}_{\phantom{\rho}\mbox{\tiny $\hat{\alpha}$}}
	   \mathfrak{g}^{\mbox{\tiny $\hat{\alpha}\hat{\beta}$}}
	    \left(\partial_{\mbox{\tiny $\hat{\alpha}$}}\phi\right)
	   \left(\partial_{\mbox{\tiny $\hat{\beta}$}}\theta\right)
	   -\frac{2}{u_{\mbox{\tiny $p$}}^2}\cot{\theta},
 \end{split}
\end{equation}
where the dilaton $\phi$ has been rescaled by
\begin{equation}\eqlabel{dilres}
\phi\:\longrightarrow\:\frac{(p-3)(7-p)}{2(5-p)}\phi
\end{equation}
in order to make the dependence on $p-3$ explicit. As mentioned earlier, the dilaton term has a trivial 
profile  for $p=2,\,3$, which implies that, in a neighbourhood of the boundary, the scalar $\theta$ exhibits 
the behaviour of a massive free scalar propagating in $AdS_{\mbox{\tiny p+1}}$ up to order $\rho^{\Delta}$.
%For $p=4$, the third term in the equation of motion $\eqref{aActEom}$ is non-zero and it
%has the effect to lower the order up to which $\theta$ behaves as a massive free scalar. 
As we will show later, in these cases the scalar $\theta(\rho)$ turns out to be tachyonic and the 
Breintelhoner-Freedman bound \cite{Breitenlohner:1982bm,Mezincescu:1984ev} is satisfied. For $p\,\neq\,2,\,3$,
the dilatonic term of \eqref{aActEom} is relevant, but it does not spoil the free massive scalar behaviour
of $\theta$.

For the sake of generality, let us rewrite the metric \eqref{FGmetric} in the following form
\begin{equation}\eqlabel{FGmetric2}
 \mathfrak{g}_{\mbox{\tiny $\hat{\alpha}\hat{\beta}$}}
  dx^{\mbox{\tiny $\hat{\alpha}$}}dx^{\mbox{\tiny $\hat{\beta}$}}\:=\:
  \frac{\mathtt{g}_{\alpha\beta}dx^{\alpha}dx^{\beta}+d\rho^2}{\rho^2}.
\end{equation}
The case of interest can be easily recovered by setting $\mathtt{g}_{\alpha\beta}\,=\,\delta_{\alpha\beta}$.
Assuming that a power expansion is valid, the most general form for the scalar $\theta$ near the boundary is
\begin{equation}\eqlabel{aBd}
 \theta\left(x,\rho\right)\:=\:
  \frac{\pi}{2}+\hat{\theta}_{1}\left(x,\rho\right),
\end{equation}
where the function $\hat{\theta}_{1}\left(x,\rho\right)$ is defined as
\begin{equation}\eqlabel{aBd1}
 \hat{\theta}_{1}\left(x,\rho\right)\:=\:
   \rho^{\alpha}\theta_{1}\left(x,\rho\right)\:\equiv\:
   \rho^{\alpha}
   \left[
   \sum_{i=0}^{\infty}\vartheta_{\alpha i}(x)\rho^{\beta_{i}}+
   \sum_{i=0}^{\infty}\psi_{\alpha i}(x)\rho^{\beta_{i}}\log{\rho}+
   \sum_{i=0}^{\infty}\sum_{j=2}^{s}\sigma_{\beta_{i},j}(x)\rho^{\beta_{i}}\log^{j}(\rho)
   \right]
\end{equation}
It's easy to notice that, up to order $\mathcal{O}\left(\rho^{3\alpha}\right)$, the equation of motion \eqref{aActEom} reduces to
\begin{equation}\eqlabel{aGen}
 0\:=\:\Box\hat{\theta}_{1}+\frac{2}{u_{\mbox{\tiny $p$}}^2}\hat{\theta}_{1}+
       \frac{\left(p-2\right)\left(p-3\right)}{5-p}\mathfrak{g}^{\mbox{\tiny $\tilde{\alpha}\tilde{\beta}$}}
        \left(\partial_{\mbox{\tiny $\tilde{\alpha}$}}\phi\right)
        \left(\partial_{\mbox{\tiny $\tilde{\beta}$}}\hat{\theta}_{1}\right)\:+\:
        \mathcal{O}\left(\rho^{3\alpha}\right).
\end{equation}
For the cases $p\,=\,2,\,3$, it's easy to see that, indeed up to order 
$\mathcal{O}\left(\rho^{3\alpha}\right)$, the equation of motion \eqref{aGen} reduces to the equation
of motion for a free massive scalar particle propagating in a background described by the metric 
$\mathfrak{g}_{\mbox{\tiny $\tilde{\alpha}\tilde{\beta}$}}$ 
(which becomes $AdS_{\mbox{\tiny $p+1$}}$ once one sets $\mathtt{g}_{\alpha\beta}\,=\,\delta_{\alpha\beta}$):
\begin{equation}\eqlabel{aFree}
 0\:=\:\Box\hat{\theta}_{1}+\frac{2}{u_{\mbox{\tiny $p$}}^2}\hat{\theta}_{1}\,+\,
       \mathcal{O}\left(\rho^{3\alpha}\right)
\qquad p\:=\:2,\,3,
\end{equation}
the mass $M$ of the scalar particle being
\begin{equation}\eqlabel{aMass}
 M^{2}\:=\:-\frac{2}{u_{p}^2}\:\equiv\:-\frac{8}{\left(5-p\right)^2}.
\end{equation} 
The $AdS$/CFT correspondence relates the mass \eqref{aMass} to the dimension
of the dual boundary operator $\mathcal{O}_{\theta}$ by
\begin{equation}\eqlabel{aMass2}
 M^{2}\:=\:\Delta\left(\Delta-p\right),
\end{equation}
and the solution for $\hat{\theta}_{1}$ can be written as
\begin{equation}\eqlabel{aSol1}
 \hat{\theta}_{1}\:=\:
    \rho^{\Delta_{-}}\left(\vartheta_{0}(x)\,+\,\ldots\right)+
    \rho^{\Delta_{+}}\left(\vartheta_{\mbox{\tiny $\left(\Delta_{+}-\Delta_{-}\right)$}}(x)\,+\,\ldots\right),
\end{equation}
where
\begin{equation}\eqlabel{ConfDim}
 \Delta_{\pm}\:=\:\frac{p}{2}\pm\frac{1}{2}\sqrt{p^2-\frac{32}{\left(5-p\right)^2}},
\end{equation}
and the conformal dimension of $\mathcal{O}_{\theta}$ is $\Delta=\Delta_{+}$.
The expression for the squared mass \eqref{aMass} clearly implies that the scalar $\theta$ is a tachyon. 
It's easy to check that the Breitenlohner-Freedman bound is satisfied for the cases of interests:
the system is then stable and the power $\alpha$ in the boundary expansion \eqref{aBd} turns out to be 
$\alpha\,=\,\Delta_{-}$\footnote{More precisely, inserting the leading order of \eqref{aBd} in \eqref{aFree}
and $\alpha$ satisfies the equation \eqref{aMass2}. The solutions $\alpha=\Delta_{-}$ and 
$\alpha=\Delta_{+}$ are the powers of $\rho$ for the non-normalizable and normalizable modes respectively.}.

For $p\,\neq\,2,\,3$, the equation \eqref{aGen} at the leading order reduces to
\begin{equation}\eqlabel{Malpha}
 M^{2}\:=\:\alpha\left(\alpha-q_{p}\right).
\end{equation}
Notice that it has the same structure of \eqref{aMass2} and can be obtained from it by a simple shift
\begin{equation}\eqlabel{pshift}
 p\:\longrightarrow\:p+\frac{\left(p-2\right)\left(p-3\right)}{5-p}=\frac{6}{5-p}\:\equiv\:q_{p}.
\end{equation}
The non-normalizable and normalizable modes therefore are
\begin{equation}\eqlabel{aSol2}
 \hat{\theta}_{1}(x,\rho)\:=\:
  \rho^{\alpha_{-}}\left(\vartheta_{0}(x)\,+\,\ldots\right)+
  \rho^{\alpha_{+}}\left(\vartheta_{\mbox{\tiny $\left(\alpha_{+}-\alpha_{-}\right)$}}(x)\,+\,\ldots\right),
\end{equation}
with
\begin{equation}\eqlabel{alphas}
 \alpha\,\equiv\,\alpha_{-}\:=\:\frac{2}{5-p},\qquad \alpha_{+}\:=\:\frac{4}{5-p}.
\end{equation}

It is possible to observe from \eqref{aActEom} by a simple counting that the terms in the boundary expansion 
of $\theta$ contributes to the divergences of the action \eqref{aActEom} up to the order 
$\mathcal{O}\left(\rho^{\alpha+\beta_{i}}\right)$, with $\alpha+\beta_{i}\,<\,(11-p)/2(5-p)$:
\begin{equation}\eqlabel{aActDiv}
 \begin{split}
  S_{\mbox{\tiny D$\left(p+2\right)$}}^{\mbox{\tiny $\theta$}}\:=\:&M\,T_{\mbox{\tiny D$\left(p+2\right)$}}\,
     \hat{\mathcal{N}}_{\mbox{\tiny $p$}}\,\int dt\,d^{\mbox{\tiny $p-1$}}x\,d\rho\:
     \frac{e^{\frac{\left(p-2\right)\left(p-3\right)}{5-p}\kappa\left(x,\rho\right)}}{
      \rho^{\frac{11-p}{5-p}}}\sqrt{\mathtt{g}}\,\times\\
   &\times\left\{1+\frac{1}{2}\left[u_{p}^2\left(\partial\hat{\theta}_{1}\right)^2-2\hat{\theta}_{1}^2\right]+
    \frac{\hat{\theta}_{1}^4}{3}-\frac{u_{p}^2}{2}\hat{\theta}_{1}^2\left(\partial\hat{\theta}_{1}\right)^2
    -\frac{u_{p}^4}{8}\left[\left(\partial\hat{\theta}_{1}\right)^2\right]^2+
    %\mathcal{O}\left(\rho^{\frac{12}{5-p}}\right)
    \ldots
    \right\}.
 \end{split}
\end{equation}
As a consequence, those terms which receive contribution from the second term in the equation of motion 
\eqref{aActEom} contribute to the finite part of the action and, therefore, the solution of equation
\eqref{aGen} provides all the terms of interest.

%Inserting the expansion 
%\eqref{aBd} in the equation of motion \eqref{aActEom}, at the leading order we get
%\begin{equation}\eqlabel{aEomL}
% \begin{split}
%  0\:&=\:%\rho^{2}\dot{\Box}\theta_{1}^{\mbox{\tiny $(0)$}}+
%         \rho^{2}\partial_{\rho}^{2}\theta_{1}^{\mbox{\tiny $(0)$}}
%         -\left(p-3\right)\rho\partial_{\rho}\theta_{1}^{\mbox{\tiny $(0)$}}-\left[M^{2}
%         -\Delta_{+}\left(\Delta_{+}-p\right)\right]\theta_{1}^{\mbox{\tiny $(0)$}}
%	 +\ldots\:=\:\\
%     &=\:\left[
%          %\left[\Delta_{+}\left(\Delta_{+}-p\right)-M^2\right]\vartheta_{0}
%	  -\left(2\Delta-p\right)\psi_{0}+2\sigma_{0,2}
%	 \right]
%         +2\left[
%          %\left[\Delta_{+}\left(\Delta_{+}-p\right)-M^2\right]\psi_{0}
%	  (2\Delta-p)\sigma_{02}+3\sigma_{03}
%	 \right]\log{\rho}+\\
%     &\hspace{.5cm}+\:\sum_{j=2}^{s-2}
%           \left(j+1\right)\left[
%	       %\left[\Delta_{+}\left(\Delta_{+}-p\right)-M^2\right]\sigma_{0,j}
%             -\left(2\Delta-p\right)\sigma_{0,\left(j+1\right)}+\left(j+2\right)\sigma_{0,\left(j+2\right)}
%	  \right]\log^{j}{\rho}+\\
%     &\hspace{.5cm}+\:s
         %\left[\Delta_{+}\left(\Delta_{+}-p\right)-M^2\right]\sigma_{0,\left(s-1\right)}
%          \left(2\Delta-p\right)\sigma_{0,s}\log^{s-1}{\rho}%+\\
%     %&+\:2^s \left[\Delta_{+}\left(\Delta_{+}-p\right)-M^2\right]\sigma_{0,s}\log^{s}{\rho}
%      +\ldots.
% \end{split}
%\end{equation}
%It is straightforward to notice that, at leading order, $\sigma_{0,j}\,=\,0\quad\forall\,j\,\ge\,2$. 
At leading order, the equation of motion for $\theta$ can be again written as the equation of motion for
a free massive scalar and the effect of the dilatonic term is just to mimic a shift of the dimensions of the 
defect as in \eqref{pshift}. As we observed earlier, the terms of the boundary expansion of $\theta$, which
contributes to the divergent part of the world-volume action, can be obtained by just solving \eqref{aGen}.
This equation can actually be reduced to the equation of a free massive scalar propagating in an $AdS$
background with fractional dimensions. %Similarly to \cite{Kanitscheider:2009as}, 
One can think to reduce the problem to D-branes wrapping an $AdS_{q+1}\times S^2$ space-time: 
the mode $\theta$ behaves as a massive free scalar propagating in an $AdS_{q+1}$ geometry and,
similarly to \cite{Kanitscheider:2009as}, the counterterms for the holographic renormalization can be
simply obtained by performing a Kaluza-Klein reduction on a $T^{q-p}$ and then analytically continuing $q$ to
take the fractional value $q_{p}$. 
From an $AdS_{q+1}$ perspective, the dual boundary operator $\mathcal{O}_{\theta}$ has conformal dimension
\begin{equation}\eqlabel{ConfDim2}
 \Delta_{q_{\mbox{\tiny $p$}}}\:=\:\frac{q_{p}}{2}+\frac{1}{2}\sqrt{q_{p}^{2}+4M^{2}},
\end{equation}
and the Breitenlohner-Freedman bound
\begin{equation}\eqlabel{BF}
 q_{p}^{2}\:\ge\:-4M^{2}\:=\:\frac{32}{\left(5-p\right)^2}.
\end{equation}
is satisfied for any $p$. Notice that $q$ ($q_{\mbox{\tiny $p$}}$) has the natural interpretation of
the number of dimensions of the boundary of the asymptotically $AdS$ space.

Furthermore, up to the order of interest, the expansion \eqref{aBd1} satisfies the equation of motion 
if and only if all the coefficients for the logarithms with high powers are zero: %$\psi_{i}\left(x\right)=0$, 
$\sigma_{i,\,j}\left(x\right)=0$ $\forall\:i\in[2/(5-p),\,4/(5-p)],\;j\in[0,\,s]$. We can write
the explicit expression for the solution
\begin{equation}\eqlabel{thetasol}
 \theta_{1}\left(x,\,\rho\right)\:=\:\rho^{\frac{2}{5-p}}\left[\vartheta_{0}\left(x\right)+
   \rho^{\frac{2}{5-p}}\left(\vartheta_{\mbox{\tiny $\left(\frac{2}{5-p}\right)$}}\left(x\right)+
   \psi_{\mbox{\tiny $\left(\frac{2}{5-p}\right)$}}\left(x\right)\log{\rho}\right)\right]+
    \mathcal{O}\left(\rho^{\frac{6}{7-p}}\right),
\end{equation}
with logarithms appearing at higher order and these higher orders not contributing to the divergence
of the action. We can check that the expansion \eqref{thetasol} coincides with the result of 
\cite{Karch:2005ms} when the D$3$/D$5$ system is considered\footnote{One can straightforwardly compare our 
expansion \eqref{thetasol}, with
equations (3.6) and (3.7) of \cite{Karch:2005ms} setting the parameters $m$ and $n$ to $4$ and $2$ 
respectively.}.

The equation of motion also tells us that the function $\psi_{\mbox{\tiny $\frac{2}{5-p}$}}(x)$ is 
non-zero only for $p=4$ and it is determined in terms of $\vartheta_{0}(x)$, 
$\mathtt{g}_{\mbox{\tiny $(0)$}}(x)$, $\kappa_{\mbox{\tiny $(0)$}}(x)$ and $\kappa_{\mbox{\tiny $(2)$}}(x)$
\begin{equation}\eqlabel{psi2}
 %\begin{split}
  %\frac{2}{5-p}
  %\left[
  % 1-\delta_{\mbox{\tiny $p,4$}}\gamma\left(p-2\right)\left(p-3\right)\kappa_{\mbox{\tiny $(2)$}}
  %\right]
  \psi_{\mbox{\tiny $\left(\frac{2}{5-p}\right)$}}\:=\:
  -\delta_{\mbox{\tiny $p,4$}}\frac{5-p}{2}
  \left[
   \dot{\Box}_{\mbox{\tiny $(0)$}}\vartheta_{\mbox{\tiny $(0)$}}+
   \frac{(p-2)(p-3)}{5-p}\gamma
   \left[
    \mathtt{g}_{\mbox{\tiny $0$}}^{\alpha\beta}
    \left(\partial_{\alpha}\kappa_{\mbox{\tiny $(0)$}}\right)
    \left(\partial_{\beta}\vartheta_{\mbox{\tiny $(0)$}}\right)+
    \frac{4}{5-p}\kappa_{\mbox{\tiny $(2)$}}
    %\left(
     \vartheta_{\mbox{\tiny $(0)$}}
     %+2\vartheta_{\mbox{\tiny $\left(\frac{2}{5-p}\right)$}}
    %\right)
   \right]
  \right],
 %\end{split}
\end{equation}
where $\dot{\Box}_{\left(0\right)}$ denotes the dalambertian operator defined through the metric 
$\mathtt{g}_{\alpha\beta}^{\mbox{\tiny $(0)$}}(x)$

\subsection{Counterterms}\label{AngEmb_Ct}

In the previous section we showed that the equation of motion for the embedding function $\theta$ is
satisfied, up to order $\mathcal{O}\left(r^{\Delta}\right)$, by a solution for a free massive scalar.
%which propagates in a $\left(q_{\mbox{\tiny $p$}}+1\right)$-dimensional $AdS$-space. 
The divergences in the action should therefore be the ones of a free scalar, which can be easily checked
from \eqref{aActDiv}.

Now, we claim that, in a similar fashion of \cite{Kanitscheider:2009as}, the counterterms for the holographic
renormalization can be computed by reducing the problem to the study of D-branes which wrap an 
$AdS_{q+1}\times S^2$, so that the counterterms are the same as the ones needed for a massive scalar 
propagating in $AdS_{q+1}$. Then, one can perform a Kaluza-Klein reduction on a $T^{q-p}$ and finally
analytically continue $q$ to the fractional value $q_{\mbox{\tiny $p$}}$.

Let us illustrate the precedure in more detail. Consider the DBI action
\begin{equation}\eqlabel{GenDBI}
 S_{\mbox{\tiny DBI}}\:=\:M\,T_{q}\,\hat{\mathcal{N}}_{q}\int dt\,d^{q-1}x\,d\rho\:\sin^2{\theta}
     \sqrt{\mathfrak{g}_{\left(q+1\right)}}\sqrt{1+u_{p}^2\left(\partial\theta\right)^2},
\end{equation}
with $\mathfrak{g}_{\mbox{\tiny $\left(q+1\right)$}}$ being the determinant of the $AdS_{q+1}$ metric.
Let us now compactify the $AdS_{q+1}$ space-time on a $T^{q-p}$:
\begin{equation}\eqlabel{Tqp}
 ds_{q+3}^2\:\equiv\:
     \mathfrak{g}_{\mbox{\tiny $AB$}}^{\mbox{\tiny $(q+1)$}}dx^{\mbox{\tiny $A$}}dx^{\mbox{\tiny $B$}}\:=\:
     \mathfrak{g}_{\mbox{\tiny $\alpha\beta$}}^{\mbox{\tiny $(p+1)$}}dx^{\mbox{\tiny $\alpha$}}
      dx^{\mbox{\tiny $\beta$}}+ 
     e^{2\frac{(p-2)(p-3)}{(5-p)(q-p)}\phi}\delta_{ab}dx^{a}dx^{b},
\end{equation}
where $\phi$ depends on the coordinates 
$\left\{x^{\alpha}\right\}_{\mbox{\tiny $\alpha=0$}}^{\mbox{\tiny $\phantom{\alpha=}p$}}$ only.
With such an ansatz, the DBI-action \eqref{GenDBI} becomes
\begin{equation}\eqlabel{GenDBI2}
  S_{\mbox{\tiny DBI}}\:=\:M\,T_{q}\,\hat{\mathcal{N}}_{q}\left(2\pi R_{\mbox{\tiny $T$}}\right)^{q-p}
     \int dt\,d^{p-1}x\,d\rho\:e^{\frac{(p-2)(p-3)}{(5-p)}\phi}\,\sin^2{\theta}
      \sqrt{\mathfrak{g}_{\left(p+1\right)}}\sqrt{1+u_{p}^2\left(\partial\theta\right)^2},
\end{equation}
which is equivalent to \eqref{aActEom}, with
\begin{equation}\eqlabel{DBIcoeff}
 T_{q}\,\hat{\mathcal{N}}_{q}\left(2\pi R_{\mbox{\tiny $T$}}\right)^{q-p}\:=\:
  T_{\mbox{\tiny D$(p+2)$}}\,\hat{\mathcal{N}}_{p}.
\end{equation}
The solution \eqref{thetasol} is indeed solution for the equation of motion from \eqref{GenDBI2}. One can 
infer that, as anticipated, the counterterms are the same needed for a massive scalar propagating in 
$AdS_{q+1}$:
\begin{equation}\eqlabel{aActCt2}
 \begin{split}
  S_{\mbox{\tiny ct}}^{\mbox{\tiny $(q)$}}\:=&\:M\,T_{q}\,\hat{\mathcal{N}}_{\mbox{\tiny $q$}}\,
      \int dt\,d^{\mbox{\tiny $q-1$}}x\,\sqrt{\gamma|_{\mbox{\tiny $\epsilon$}}}
      e^{\frac{\left(p-2\right)\left(p-3\right)}{5-p}\phi\left(x,\epsilon\right)}
      \left\{
       -\frac{1}{q}+
     \left[\frac{\delta_{q,2}}{4}\log{\epsilon}-\frac{\varepsilon_{q,2}}{2q(q-1)(q-2)}\right]R_{\gamma}^{(q)}+
      \right.\\
     &%\left.
       +\frac{\delta_{q,4}}{32}\log{\epsilon}
        \left(R_{\gamma}^{\mbox{\tiny $AB$}}R^{\gamma}_{\mbox{\tiny $AB$}}-\frac{1}{3}R^{2}_{\gamma}\right)
       +\left[\frac{q-\Delta}{2}+\frac{\delta_{\mbox{\tiny $q,2\Delta$}}}{2\log{\epsilon}}
       +\frac{q-\Delta}{2\left(q-1\right)}
        \left(-\frac{\delta_{\mbox{\tiny $q,2\Delta-2$}}}{2}\log{\epsilon}\right.
      \right.\\
     &\left.\left.+\frac{\varepsilon_{\mbox{\tiny $q,2\Delta-2$}}}{2\left(2\Delta-q-2\right)}\right)
        R_{\gamma}\right]u_{\mbox{$p$}}^2\hat{\theta}_{\mbox{\tiny $1$}}\left(x,\epsilon\right)^2+\\
     &+\left.\left[
        \frac{\varepsilon_{\mbox{\tiny $q,2\Delta-2$}}}{2\left(2\Delta-q-2\right)}-
        \frac{\delta_{q,2\Delta-2}}{2}\log{\epsilon}
       \right]
        u_{\mbox{\tiny $p$}}^2\hat{\theta}_{\mbox{\tiny $1$}}(x,\epsilon)
         \Box_{\gamma}^{\mbox{\tiny $\left(q\right)$}}\hat{\theta}_{\mbox{\tiny $1$}}(x,\epsilon)
      \right\},
 \end{split}
\end{equation}
where $\varepsilon_{p,l}$ is $0$ if $p=l$ and $1$ otherwise. Compactifying on the $T^{q-p}$, 
%the Ricci tensor $R_{\mbox{\tiny $AB$}}^{\gamma_{\mbox{\tiny $(q)$}}}$, 
the curvature scalar $R_{\gamma}^{\mbox{\tiny $\left(q\right)$}}$ and the dalambertian operator  
$\Box_{\gamma}^{\mbox{\tiny $\left(q\right)$}}$ reduce to
\begin{equation}\eqlabel{Rbox}
 \begin{split}
  %&R_{\mbox{\tiny $\alpha\beta$}}^{\gamma_{\mbox{\tiny $(q)$}}}\:=\:
  % R_{\mbox{\tiny $\alpha\beta$}}^{\gamma_{\mbox{\tiny $(p)$}}}
  % -\frac{\left(p-2\right)\left(p-3\right)}{\left(5-p\right)}\nabla_{\alpha}\nabla_{\beta}\phi+\\
  %&\phantom{R_{\mbox{\tiny $\alpha\beta$}}^{\gamma_{\mbox{\tiny $(q)$}}}\:=\:
  %  R_{\mbox{\tiny $\alpha\beta$}}^{\gamma_{\mbox{\tiny $(p)$}}}}
  % +\left(p+q-4\right)\frac{\left(p-2\right)^2\left(p-3\right)^2}{\left(5-p\right)^2\left(q-p\right)^2}
  %  \left[
  %   \left(\partial_{\alpha}\phi\right)\left(\partial_{\beta}\phi\right)-
  %   \gamma_{\alpha\beta}^{\mbox{\tiny $(p)$}}\left(\partial\phi\right)^2
  %  \right]\\
  &R_{\gamma}^{\mbox{\tiny $\left(q\right)$}}\:=\:R_{\gamma}^{\mbox{\tiny $\left(p\right)$}}
   -2\frac{\left(p-2\right)\left(p-3\right)}{5-p}\Box_{\mbox{\tiny $p$}}\phi
  %&\phantom{R_{\gamma}^{\mbox{\tiny $\left(q\right)$}}\:=\:R_{\gamma}^{\mbox{\tiny $\left(p\right)$}}}
    -\frac{\left(p-2\right)^2\left(p-3\right)^2}{\left(5-p\right)^2}
    %\left[\left(q-2\right)\left(q-1\right)+\left(p-1\right)\left(p-2\right)\right]
    \frac{q-p+1}{q-p}
     \left(\partial\phi\right)^2\\
  &\Box_{\gamma}^{\mbox{\tiny $\left(q\right)$}}\:=\:
   \Box_{\gamma}^{\mbox{\tiny $\left(p\right)$}}+\frac{\left(p-2\right)\left(p-3\right)}{5-p}
    \gamma^{\mbox{\tiny $\hat{\alpha}\hat{\beta}$}}
    \left(\partial_{\mbox{\tiny $\hat{\alpha}$}}\phi\right)\partial_{\hat{\beta}}
 \end{split}
\end{equation}
Inserting \eqref{Rbox} in \eqref{aActCt2} and analytically continuing $q$ to $q_{p}$ (the explicit
expression of $q_{\mbox{\tiny $p$}}$ is in \eqref{pshift}), we obtain the counterterms we were looking for.
Notice that, once the analytic continuation to $q_{\mbox{\tiny $p$}}$ is performed, one can realize that 
the first two terms in the second line of \eqref{aActCt2} do not contribute for any $p$ given that 
$\delta_{\mbox{\tiny $q$},4}$ would be satisfied just for $p=7/2$. Similarly, the term proportional to
$\left(\log{\epsilon}\right)^{-1}$ does not appear for any values of $p$ of interest (the constraint
$q-2\Delta=0$ is never satisfied). The counterterm action can be therefore written as
\begin{equation}\eqlabel{aActCt3}
 \begin{split}
  S_{\mbox{\tiny ct}}^{\mbox{\tiny $(\theta)$}}\:=&\:
      M\,T_{\mbox{\tiny D$(p+2)$}}\,\hat{\mathcal{N}}_{\mbox{\tiny $p$}}\,
      \int dt\,d^{\mbox{\tiny $p-1$}}x\,\sqrt{\gamma|_{\mbox{\tiny $\epsilon$}}}
      e^{\frac{\left(p-2\right)\left(p-3\right)}{5-p}\phi\left(x,\epsilon\right)}\times\\
     &\times
      \left\{
       -\frac{5-p}{6}+
     \left[\frac{\delta_{p,2}}{4}\log{\epsilon}-\varepsilon_{p,2}\frac{(5-p)^3}{24(p+1)(p-2)}\right]
     \left[
      R_{\gamma}-2\frac{(p-2)(p-3)}{5-p}\nabla^2\phi-\right.\right.\\
     &\left.
      -\frac{(p-2)(p-3)(p^2-6p+11)}{(5-p)^2}\left(\partial\phi\right)^2
     \right]+
      \left[\frac{1}{5-p}
       +\frac{1}{p+1}
        \left(-\frac{\delta_{\mbox{\tiny $p,4$}}}{2}\log{\epsilon}+
        \frac{\varepsilon_{\mbox{\tiny $p,4$}}}{4}\frac{5-p}{p-4}\right)\times\right.\\
       &\left.\times\left(
        R_{\gamma}-2\frac{(p-2)(p-3)}{5-p}\nabla^2\phi
        -\frac{(p-2)(p-3)(p^2-6p+11)}{(5-p)^2}\left(\partial\phi\right)^2
       \right)
       \right]u_{\mbox{\tiny $p$}}^2\hat{\theta}_{\mbox{\tiny $1$}}\left(x,\epsilon\right)^2+\\
       &\left.+\left[
        \frac{\varepsilon_{\mbox{\tiny $p,4$}}}{4}\frac{5-p}{p-4}-
        \frac{\delta_{p,4}}{2}\log{\epsilon}
       \right]
        u_{\mbox{\tiny $p$}}^{2}\hat{\theta}_{\mbox{\tiny $1$}}(x,\epsilon)
        \left[
         \Box_{\gamma}^{\mbox{\tiny $\left(p\right)$}}\theta(x,\epsilon)+
         \frac{(p-2)(p-3)}{5-p}\gamma^{\alpha\beta}\left(\partial_{\alpha}\phi\right)
          \left(\partial_{\beta}\hat{\theta}_{\mbox{\tiny $1$}}\right)
        \right]
      \right\},
 \end{split}
\end{equation}
It is easy to notice that for $p=3$, the counterterm action \eqref{aActCt3} coincides with the one
discussed in \cite{Karch:2005ms}.

%The most general expression for the counterterm action which renormalizes the divergences \eqref{aActDiv}
%is
%\begin{equation}\eqlabel{aActCt}
% \begin{split}
%  S_{\mbox{\tiny ct}}^{\mbox{\tiny $\theta$}}\:=&\:M\,T_{\mbox{\tiny D$\left(p+2\right)$}}\,
%      \hat{\mathcal{N}}_{\mbox{\tiny $p$}}\,\int dt\,d^{\mbox{\tiny $p-1$}}x\,
%      e^{\frac{\left(p-2\right)\left(p-3\right)}{5-p}\phi}\sqrt{\gamma|_{\mbox{\tiny $\epsilon$}}}
%      \left[
%       \alpha_{1}+\alpha_{2}R_{\gamma}+\alpha_{3}\left(\partial_{\alpha}\phi\right)^{2}+\right.\\
%       &\left.+\left(\alpha_{4}+\frac{\alpha_{5}}{\log{\epsilon}}+\alpha_{6}R_{\gamma}\right)
%        \theta\left(x,\epsilon\right)^2+
%       \alpha_{7}\theta(x,\epsilon)\Box_{\gamma}\theta(x,\epsilon)
%      \right],
% \end{split}
%\end{equation}
%where the coefficients $\alpha_{i}$ can be easily found to be
%\begin{equation}\eqlabel{coeffs}
% \alpha_{1}\:=\:-\frac{1}{q_{\mbox{\tiny $p$}}}
%\end{equation}

\subsection{One-point correlator}\label{AngEmb_1pt}

We can use the holographic renormalized action obtained in the previous section to compute the 
one-point correlator for the operator $\mathcal{O}_{\mbox{\tiny $\theta$}}$ dual to the mode 
$\theta$ which describes the embedding of the D$(p+2)$-branes. The standard $AdS/CFT$ correspondence 
prescribes the one-point correlator to be
\begin{equation}\eqlabel{1ptcorr}
 \langle\mathcal{O}_{\mbox{\tiny $\theta$}}\rangle_{\mbox{\tiny $(q)$}}\:=\:
  \lim_{\epsilon\rightarrow0}\frac{1}{\epsilon^{\Delta}}\frac{1}{\sqrt{\gamma|_{\mbox{\tiny $\epsilon$}}}}
   \frac{\delta S_{\mbox{\tiny DBI}}^{\mbox{\tiny $\left(\theta\right)$}}\Big|_{\mbox{\tiny ren}}}{
    \delta\theta(x,\epsilon)}.
\end{equation}
For the case of branes embedded in a background generated by a stack of D$p$-branes, the prescription 
\eqref{1ptcorr} can be still used, with a little subtlety. As we showed in the previous section, one
can map the problem to the study of branes which wrap an $AdS_{q+1}\times S^{2}$ subspace. At this level,
one can apply the standard prescription \eqref{1ptcorr} and then consider the metric ansatz 
\eqref{Tqp}. The final step is the analytic continuation of $q$ to $q_{p}$. Considering also the relation
\eqref{GenDBI2} between the action in $(q+1)$-dimensional asymptotically $AdS$ space \eqref{GenDBI} and
the original one \eqref{aActEom}, one can infer the relation between the one-point correlator 
\eqref{1ptcorr} and the one in $p$-dimensions
\begin{equation}\eqlabel{1ptcorr2}
 \langle\mathcal{O}_{\mbox{\tiny $\theta$}}\rangle_{\mbox{\tiny $(p)$}}\:=\:
 e^{2\frac{p-2}{5-p}\left(\delta_{\mbox{\tiny $p,3$}}+\frac{7-p}{p-3}\varepsilon_{\mbox{\tiny $p,3$}}\right)
   \kappa_{\mbox{\tiny $(0)$}}}\left(2\pi R_{\mbox{\tiny $T$}}\right)^{q-p}
  \langle\mathcal{O}_{\mbox{\tiny $\theta$}}\rangle_{\mbox{\tiny $(q)$}}
\end{equation}

Let us analyze this procedure in detail. 
Applying the prescription \eqref{1ptcorr} to the action
\eqref{GenDBI}, whose counterterms are \eqref{aActCt2}, one obtain:
\begin{equation}\eqlabel{1ptcorrGen}
  \langle\mathcal{O}_{\mbox{\tiny $\theta$}}\rangle_{\mbox{\tiny $(q)$}}\:=\:
   \lim_{\epsilon\rightarrow0}\frac{1}{\epsilon^{\frac{4}{5-p}}}
    \frac{\epsilon^{p}\,e^{-\frac{(p-2)(p-3)}{5-p}\phi(x,\epsilon)}}{\sqrt{\mathtt{g}(x,\epsilon)}}
    \left[
     \frac{\delta S_{\mbox{\tiny DBI}}^{
      \mbox{\tiny $\left(\theta\right)$}}\Big|_{\mbox{\tiny on-shell}}}{
      \delta\theta(x,\epsilon)}+
     \frac{\delta S_{\mbox{\tiny DBI}}^{
      \mbox{\tiny $\left(\theta\right)$}}\Big|_{\mbox{\tiny ct}}}{
      \delta\theta(x,\epsilon)}
    \right].
\end{equation}
Let us focus on the first term of \eqref{1ptcorrGen}, whose explicit expression is
\begin{equation}\eqlabel{1ptcorrGen1}
  \begin{split}
   \frac{1}{\epsilon^{\frac{4}{5-p}}}
    &\frac{\epsilon^{p}\,e^{-\frac{(p-2)(p-3)}{5-p}\phi(x,\epsilon)}}{\sqrt{\mathtt{g}(x,\epsilon)}}
     \frac{\delta S_{\mbox{\tiny DBI}}^{
      \mbox{\tiny $\left(\theta\right)$}}\Big|_{\mbox{\tiny on-shell}}}{
      \delta\theta(x,\epsilon)}\:=\:
      -\frac{u_{p}^2}{\varepsilon^{\frac{2}{(5-p)}}}\frac{2}{5-p}\vartheta_{\mbox{\tiny $(0)$}}-\\
    &-u_{p}^2
     \left[
      \frac{4}{5-p}\vartheta_{\mbox{\tiny $(\frac{2}{5-p})$}}+
      \psi_{\mbox{\tiny $\left(\frac{2}{5-p}\right)$}}
     %\right.\\
     %&-\frac{5-p}{2}\delta_{\mbox{\tiny $p,4$}}
     % \left[
     %  \dot{\Box}_{\mbox{\tiny $(0)$}}\vartheta_{\mbox{\tiny $(0)$}}+
     %  \frac{(p-2)(p-3)}{5-p}\gamma
     %  \left[
     %   \mathtt{g}_{\mbox{\tiny $0$}}^{\alpha\beta}
     %   \left(\partial_{\alpha}\kappa_{\mbox{\tiny $(0)$}}\right)
     %   \left(\partial_{\beta}\vartheta_{\mbox{\tiny $(0)$}}\right)+
     %   \frac{4}{5-p}\kappa_{\mbox{\tiny $(2)$}}
     %   \vartheta_{\mbox{\tiny $(0)$}}
     % \right]
     %\right]\times\\
    %&\left.\times
     \left(1+\frac{4}{5-p}\log{\varepsilon}\right)
     \right]+\mathcal{O}\left(\rho^{\frac{2}{5-p}}\right).
   \end{split}
\end{equation}
There are only two type of divergences: one of order 
$\mathcal{O}\left(\epsilon^{-\frac{2}{5-p}}\right)$ and the other one is a logarithm which is
present just for the case of $p=4$. From \eqref{aActCt2}, it is straightforward to notice that only
the terms $\theta^{2}$ and $\theta\Box_{\gamma}^{\mbox{\tiny $(q)$}}\theta$ contribute to 
renormalize the correlator $\langle\mathcal{O}_{\mbox{\tiny $\theta$}}\rangle$.
Let us now consider the contribution from the counterterm action \eqref{aActCt2}:
\begin{equation}\eqlabel{1ptcorrCt}
 \begin{split}
  \frac{1}{\epsilon^{\frac{4}{5-p}}}
  &\frac{e^{-\frac{(p-2)(p-3)}{5-p}\phi(x,\epsilon)}}{\sqrt{\mathtt{g}(x,\epsilon)}}
   \frac{\delta S_{\mbox{\tiny DBI}}^{
      \mbox{\tiny $\left(\theta\right)$}}\Big|_{\mbox{\tiny ct}}}{
      \delta\theta(x,\epsilon)}\:=\:
      \frac{u_{\mbox{\tiny $p$}}^2}{\epsilon^{\frac{4}{5-p}}}
      \left[\frac{2}{5-p}
       +\frac{2}{p+1}
        \left(-\frac{\delta_{\mbox{\tiny $p,4$}}}{2}\log{\epsilon}+
        \frac{\varepsilon_{\mbox{\tiny $p,4$}}}{4}\frac{5-p}{p-4}\right)\times\right.\\
       &\left.\times\left(
        R_{\gamma}-2\frac{(p-2)(p-3)}{5-p}\Box_{\gamma}^{\mbox{\tiny $\left(p\right)$}}\phi
        -\frac{(p-2)(p-3)(p^2-6p+11)}{(5-p)^2}\left(\partial\phi\right)^2
       \right)
       \right]\hat{\theta}_{\mbox{\tiny $1$}}\left(x,\epsilon\right)+\\
       &+\left[
        \frac{\varepsilon_{\mbox{\tiny $p,4$}}}{4}\frac{5-p}{p-4}-
        \frac{\delta_{p,4}}{2}\log{\epsilon}
       \right]
        %\theta(x,\epsilon)
        \left[
         \Box_{\gamma}^{\mbox{\tiny $\left(p\right)$}}\hat{\theta}_{\mbox{\tiny $1$}}(x,\epsilon)+
         \frac{(p-2)(p-3)}{5-p}\gamma^{\alpha\beta}\left(\partial_{\alpha}\phi\right)
          \left(\partial_{\beta}\hat{\theta}_{\mbox{\tiny $1$}}\right)
        \right]
 \end{split}
\end{equation}
Notice that the terms containing the scalar curvature $R_{\gamma}$, the dalambertian operator 
$\Box_{\gamma}^{\mbox{\tiny $\left(p\right)$}}$ and the derivatives of the dilaton contribute
just for $p=4$, since the leading order of these terms is $\mathcal{O}\left(\epsilon^{\frac{2}{5-p}+2}\right)$
and therefore they contribute to the correlator at order 
$\mathcal{O}\left(\epsilon^{2\frac{4-p}{5-p}}\right)$. Since these terms are relevant for $p=4$ only, they
contributes just to cancel the logarithmic divergence which arises in \eqref{1ptcorrGen1}.

Taking into account the contributions \eqref{1ptcorrGen1} and \eqref{1ptcorrCt}, the one-point correlator
for the operator $\mathcal{O}_{\mbox{\tiny $\theta$}}$ is
\begin{equation}\eqlabel{1ptcorrFin}
 \begin{split}
 \langle\mathcal{O}_{\mbox{\tiny $\theta$}}\rangle_{\mbox{\tiny $(p)$}}\:=\:
  -&
  e^{2\frac{p-2}{5-p}\left(\delta_{\mbox{\tiny $p,3$}}+\frac{7-p}{p-3}\varepsilon_{\mbox{\tiny $p,3$}}\right)
   \kappa_{\mbox{\tiny $(0)$}}}
  %e^{\frac{(p-2)(p-3)}{5-p}\kappa_{\mbox{\tiny $(0)$}}}
  u_{\mbox{\tiny $p$}}^2
  \left\{
   \frac{2}{5-p}\vartheta_{\mbox{\tiny $\left(\frac{2}{5-p}\right)$}}
   -\delta_{\mbox{\tiny $p,4$}}\frac{5-p}{2}
   \Big[
    \dot{\Box}_{\mbox{\tiny $(0)$}}\vartheta_{\mbox{\tiny $(0)$}}+\right.\\
   &+\left.\left.\frac{(p-2)(p-3)}{5-p}\gamma
    \left[
     \mathtt{g}_{\mbox{\tiny $0$}}^{\alpha\beta}
     \left(\partial_{\alpha}\kappa_{\mbox{\tiny $(0)$}}\right)
     \left(\partial_{\beta}\vartheta_{\mbox{\tiny $(0)$}}\right)+
     \frac{4}{5-p}\kappa_{\mbox{\tiny $(2)$}}
     %\left(
      \vartheta_{\mbox{\tiny $(0)$}}
      %+2\vartheta_{\mbox{\tiny $\left(\frac{2}{5-p}\right)$}}
     %\right)
    \right]
   \right]
  \right\}
 \end{split}
\end{equation}
Notice that \eqref{1ptcorrFin} correctly reproduces the result of \cite{Karch:2005ms} for the D$3$/D$5$-system
once $p$ is set to $3$.

\section{Brane intersections with codimension-$k$ defect}\label{Codk}

In this section we extend the previous discussion to systems with a codimension-$k$ defect, with $k=0,2$.
The supersymmetric systems of interests are therefore D$p$/D$(p+4)$ (codimension-$0$ defect) and 
D$p$/D$p$ (codimension-$2$ defect). 
Generally speaking, the probe branes wrap an internal $(3-k)$-sphere $S^{3-k}$ and the action and the 
equation of motion for the embedding mode $\theta$ can be easily obtained from
\eqref{aActEom} by mapping $\sin^2{\theta}$ to $\sin^{3-k}{\theta}$ and $2\cot{\theta}/
u_{\mbox{\tiny $p$}}^2$ to $(3-k)\cot{\theta}/u_{\mbox{\tiny $p$}}^2$ respectively, and
$(p-2)\,\rightarrow\,(p-(k+1))$ in the dilaton factor of both the action and the equation of motion:
\begin{equation}\eqlabel{aActEomK}
 \begin{split}
  &S_{\mbox{\tiny D$(p+4-2k)$}}^{\mbox{\tiny $\theta$}}\:=\: 
   M\,T_{\mbox{\tiny D$(p+4-2k)$}}\hat{\mathcal{N}}_{\mbox{\tiny $\{p,k\}$}}
   \int dt\,d^{p-k}x\,\rho\:e^{\frac{(p-(k+1))(p-3)}{5-p}\phi}\sin^{3-k}{\theta}
   \sqrt{\mathfrak{g}}\sqrt{1+u_{\mbox{\tiny $p$}}^2\left(\partial\theta\right)^2}\\
  &0\:=\:\Box\theta-
     	  \frac{1}{2}\mathfrak{g}^{\mbox{\tiny $\hat{\alpha}\hat{\beta}$}}
	  \left(\partial_{\mbox{\tiny $\hat{\alpha}$}}\theta\right)
	  \frac{\partial_{\mbox{\tiny $\hat{\beta}$}}\left[1+
           u_{\mbox{\tiny $p$}}^2\left(\partial\theta\right)^2\right]}{1+
	    u_{\mbox{\tiny $p$}}^2\left(\partial\theta\right)^2}+
           \frac{(p-(k+1))(p-3)}{5-p}
	   %\delta^{\rho}_{\phantom{\rho}\mbox{\tiny $\hat{\alpha}$}}
	   \mathfrak{g}^{\mbox{\tiny $\hat{\alpha}\hat{\beta}$}}
	    \left(\partial_{\mbox{\tiny $\hat{\alpha}$}}\phi\right)
	   \left(\partial_{\mbox{\tiny $\hat{\beta}$}}\theta\right)
	   -\frac{3-k}{u_{\mbox{\tiny $p$}}^2}\cot{\theta},
 \end{split}
\end{equation}
% simply mapping $\sin^2{\theta}$ to $\sin^{3-k}{\theta}$ and $2\cot{\theta}/
%u_{\mbox{\tiny $p$}}^2$ to $(3-k)\cot{\theta}/u_{\mbox{\tiny $p$}}^2$ respectively. 
Using the same boundary expansion \eqref{aBd1} for the embedding function, the equation of motion at the 
leading order gives the squared-mass relation \eqref{Malpha}, with the replacement 
\begin{equation}\eqlabel{qpk}
q_{\mbox{\tiny $p$}}\:\rightarrow
q_{\mbox{\tiny $\{p,k\}$}}\:=\:p+\frac{(p-(k+1))(p-3)}{5-p}-k+1\:=\:2\frac{4-k}{5-p},
\end{equation}
The non-normalizable and normalizable modes are
\begin{equation}\eqlabel{aSol3}
 \hat{\theta}_{1}\:=\:\rho^{\alpha_{\mbox{\tiny $-$}}}\left(\vartheta_{\mbox{\tiny $0$}}+\ldots\right)+
   \rho^{\alpha_{\mbox{\tiny $+$}}}\left(\vartheta_{\mbox{\tiny $\left(\alpha_{+}-\alpha_{-}\right)$}}+
    \ldots\right), 
\end{equation}
with
\begin{equation}\eqlabel{alphas2}
 \alpha_{\mbox{\tiny $-$}}=\frac{2}{5-p}, \qquad
 \alpha_{\mbox{\tiny $+$}}=2\frac{3-k}{5-p}.
\end{equation}
The boundary expansion \eqref{aBd1} is constrained by the equation of motion \eqref{aActEomK} to have 
$\sigma_{i,j}(x)=0, \;\forall j\in[2,s],\;\forall i\in[\alpha_{-},\,\alpha_{+}]$.
The terms in the expansion \eqref{aBd1} contributes to the 
divergences up to the order $\mathcal{O}\left(\alpha+\beta_{i}\right)$ with 
$\alpha+\beta_{i}<\left[(13-2k)-p\right]/2(5-p)$. Up to the order of interest, there are no
higher power logarithm in the solution for the embedding mode and this coincides with the solution
for a massive free particle in a $(q_{\mbox{\tiny $\{p,k\}$}}+1)$-dimensional asymptotically $AdS$-space.
From this $AdS$-perspective, the dual operator $\mathcal{O}_{\mbox{\tiny $\theta$}}$ has conformal dimension
$\Delta=\alpha_{\mbox{\tiny $+$}}$.
The correct counterterms can be obtained again considering branes $(q+1)$-dimensional asymptotically 
$AdS$-space, for which the counterterms are given in \eqref{aActCt2} (after replacing 
$(p-2)\rightarrow(p-k-1)$ in the dilaton factor), dimensional reducing them
on a torus $T^{q-p+k-1}$, and analytically continuing $q$ to $q_{\mbox{\tiny $\{p,k\}$}}$. 
Notice that for codimension-$2$ defects, the Breitenlohner-Freedman bound is saturated
and the solution of the equation of motion acquires the following form
\begin{equation}\eqlabel{DpDpSol}
 \hat{\theta}_{\mbox{\tiny $1$}}(x,\rho)\:=\:\rho^{\frac{2}{5-p}}
  \left[\vartheta_{\mbox{\tiny $0$}}(x)+\psi_{\mbox{\tiny $0$}}(x)\log{\rho}+\ldots\right].
\end{equation}
This is the only case in which the term proportional to $(\log{\epsilon})^{-1}$ appears. 
The one-point correlator $\langle\mathcal{O}_{\mbox{\tiny $\theta$}}\rangle_{\mbox{\tiny $(p-k)$}}$ is
\begin{equation}\eqlabel{1ptcorr3}
 \begin{split}
  \langle\mathcal{O}_{\mbox{\tiny $\theta$}}\rangle_{\mbox{\tiny $(p-k)$}}\:&=\:
  % e^{\frac{(p-k-1)(p-3)}{5-p}\kappa_{\mbox{\tiny $(0)$}}}
e^{2\frac{p-k-1}{5-p}\left(\delta_{\mbox{\tiny $p,3$}}+\frac{7-p}{p-3}\varepsilon_{\mbox{\tiny $p,3$}}\right)
   \kappa_{\mbox{\tiny $(0)$}}}
   \left(2\pi R_{\mbox{\tiny $T$}}\right)^{q-p+k-1}
   \langle\mathcal{O}_{\mbox{\tiny $\theta$}}\rangle_{\mbox{\tiny $(q)$}}\:=\\
  &=\:
   %e^{\frac{(p-k-1)(p-3)}{5-p}\kappa_{\mbox{\tiny $(0)$}}}
e^{2\frac{p-k-1}{5-p}\left(\delta_{\mbox{\tiny $p,3$}}+\frac{7-p}{p-3}\varepsilon_{\mbox{\tiny $p,3$}}\right)
   \kappa_{\mbox{\tiny $(0)$}}}
   \left(2\pi R_{\mbox{\tiny $T$}}\right)^{q-p+k-1}
   \lim_{\epsilon\rightarrow0}
   \frac{\varepsilon_{\mbox{\tiny $k,2$}}+\delta_{\mbox{\tiny $k,2$}}\log{\epsilon}}{\epsilon^{\Delta}}
   \frac{1}{\sqrt{\gamma|_{\mbox{\tiny $\epsilon$}}}}
   \frac{\delta S_{\mbox{\tiny $DBI$}}^{\mbox{\tiny $(\theta)$}}\Big|_{\mbox{\tiny ren}}}{
    \delta\theta(x,\epsilon)}
 \end{split}
\end{equation}
For the supersymmetric case with $k=2$, one obtains
\begin{equation}\eqlabel{1ptcorr4}
  \langle\mathcal{O}_{\mbox{\tiny $\theta$}}\rangle_{\mbox{\tiny $(p-2)$}}\:=\:\vartheta_{\mbox{\tiny (0)}}.
\end{equation}
This means that in the D$p$/D$p$ the one-point correlator for the operator dual to the embedding mode 
is determined by the coefficient of the normalizable mode: the observation made in \cite{Karch:2005ms}
that the brane separation appears as a vev for the D$3$/D$3$-system extends to all the other systems with a 
codimension-$2$ defect ($p<5$).

\section{Conclusion}\label{Conc}

In this paper, we extended the holographic renormalization method to probe D-branes in non-conformal 
backgrounds. The key observation is that, as for theories with no flavours \cite{Kanitscheider:2009as},
the computation can be reduced to the computation of counterterms for probe branes in (higher-dimensional)
asymptotically $AdS$ space-times. More specifically, the mode which describes the embedding of the probe
branes behaves as a free massive scalar propagating in a higher-dimensional $AdS$ space-time,
at least in a neighbourhood of the boundary and for all the orders which contribute to the
divergent terms of the action. The enhancement of the number of dimensions is a direct consequence of
the presence of a non-trivial profile for the dilaton.  We explicitly showed that the DBI-action
for the probe D-branes in non-conformal backgrounds is equivalent to the DBI-action for probe branes
in a higher-dimensional $AdS$-space: the original form for the DBI-action can be recovered by
a Kaluza-Klein reduction on a warped torus $T^{q-p+k-1}$, where the warped factor depends on the dilaton
field. Strictly speaking, the extra number of dimensions is fractional, so this picture has been made
useful by considering number of dimensions of this $AdS$-space as an arbitrary integer and then performing
analytical continuation to the actual fractional value after the computation.
From this higher dimensional $AdS$ view-point, we observed that the angular embedding mode strictly 
satisfies the Breitenlohner-Freedman bound, except for D$p$/D$p$ systems for which
the bound is saturated. For the latter class of systems, the one-point correlator is expressed
in terms of the coefficient of the normalizable mode.
This perspective drastically simplifies the computation of the counterterms for the holographic
renormalization: they are just given by the counterterms for a massive scalar particle propagating
in this $AdS$-space. Furthermore, it allows to straightforwardly apply the standard $AdS/CFT$ 
prescription for the computation of one-point correlators. 

One can extend this view-point also to the simple case of the linear embedding. In section \ref{LinEmb},
we easily computed the single counterterm needed by inspecting the only divergent term of the action
in a neighbourhood of the boundary. We also argued that the only divergent term appearing in the action
has the same behaviour of the one that one would obtain in the action of branes in a
 ``$6/(5-p)+1$''-dimensional $AdS$-space. More precisely, it is easy to see that DBI-action of
branes in $AdS_{q+1}$, $q$ being again an arbitrary integer,
\begin{equation}\eqlabel{zAct2}
 S_{\mbox{\tiny DBI}}^{\mbox{\tiny $\left(z\right)$}}\:=\:
        M\,T_{\mbox{\tiny $q$}}\,\hat{\mathcal{N}}_{\mbox{\tiny $q$}}
        \int dt\,d^{q-1}x\,d\rho\,\sqrt{\mathfrak{g}_{\mbox{\tiny $(q+1)$}}}\,
	 \sqrt{1+\rho^{-2}\left(\partial z\right)^{2}}
\end{equation}
reduces to \eqref{zActEom} if a Kaluza-Klein reduction is performed on the metric ansatz \eqref{Tqp}
and then $q$ is analytically continued to $6/(5-p)$, and \eqref{bd} is the solution of the equation
of motion from \eqref{zAct2}, for 
$\mathtt{g_{\mbox{\tiny $\alpha\beta$}}}=\delta_{\alpha\beta}$.
Inserting the solution \eqref{bd} in the action \eqref{zAct2}, one can see that the only divergent
term comes from the volume of $AdS_{q+1}$, for which the counterterms are well-known 
\cite{Skenderis:2002wp}. For $\mathtt{g_{\mbox{\tiny $\alpha\beta$}}}=\delta_{\alpha\beta}$, only the
term proportional to the volume of the boundary of $AdS_{q+1}$ contributes.

Even if the simplicity of the linear description does not explicitly require any different perspective,
it is useful to have a consistent and completely general $AdS$-viewpoint.

The remarkable observation of \cite{Kanitscheider:2009as} that non-conformal backgrounds can 
be mapped into higher dimensional asymptotically $AdS$-geometries drastically simplifies the study of the 
dynamics of such systems, which may be determined in terms of the dynamics of conformal systems.
We showed that also the degrees of freedom that can be introduced by adding probe branes may behave
as degrees of freedom in asymptotically $AdS$ space-times and, therefore, it may be possible
to determine all the physics of these systems in terms of known results for the conformal case.

\section*{Acknowledgments}

It is a pleasure to thank Kasper Peeters and Simon Ross for useful discussions and comments on the manuscript.
This work is supported by STFC Rolling Grant.

\bibliographystyle{utphys}
\bibliography{gaugegravityrefs}

\providecommand{\href}[2]{#2}\begingroup\raggedright\begin{thebibliography}{10}

\bibitem{Maldacena:1997re}
J.~M. Maldacena, ``{The large N limit of superconformal field theories and
  supergravity},'' {\em Adv. Theor. Math. Phys.} {\bf 2} (1998)  231--252,
\href{http://arxiv.org/abs/hep-th/9711200}{{\tt arXiv:hep-th/9711200}}.
%%CITATION = HEP-TH/9711200;%%.

\bibitem{Gubser:1998bc}
S.~S. Gubser, I.~R. Klebanov, and A.~M. Polyakov, ``{Gauge theory correlators
  from non-critical string theory},''
  \href{http://dx.doi.org/10.1016/S0370-2693(98)00377-3}{{\em Phys. Lett.} {\bf
  B428} (1998)  105--114},
\href{http://arxiv.org/abs/hep-th/9802109}{{\tt arXiv:hep-th/9802109}}.
%%CITATION = HEP-TH/9802109;%%.

\bibitem{Witten:1998qj}
E.~Witten, ``{Anti-de Sitter space and holography},'' {\em Adv. Theor. Math.
  Phys.} {\bf 2} (1998)  253--291,
\href{http://arxiv.org/abs/hep-th/9802150}{{\tt arXiv:hep-th/9802150}}.
%%CITATION = HEP-TH/9802150;%%.

\bibitem{Aharony:1999ti}
O.~Aharony, S.~S. Gubser, J.~M. Maldacena, H.~Ooguri, and Y.~Oz, ``{Large N
  field theories, string theory and gravity},''
  \href{http://dx.doi.org/10.1016/S0370-1573(99)00083-6}{{\em Phys. Rept.} {\bf
  323} (2000)  183--386},
\href{http://arxiv.org/abs/hep-th/9905111}{{\tt arXiv:hep-th/9905111}}.
%%CITATION = HEP-TH/9905111;%%.

\bibitem{Susskind:1998dq}
L.~Susskind and E.~Witten, ``{The holographic bound in anti-de Sitter space},''
\href{http://arxiv.org/abs/hep-th/9805114}{{\tt arXiv:hep-th/9805114}}.
%%CITATION = HEP-TH/9805114;%%.

\bibitem{Henningson:1998gx}
M.~Henningson and K.~Skenderis, ``{The holographic Weyl anomaly},'' {\em JHEP}
  {\bf 07} (1998)  023,
\href{http://arxiv.org/abs/hep-th/9806087}{{\tt arXiv:hep-th/9806087}}.
%%CITATION = HEP-TH/9806087;%%.

\bibitem{Henningson:1998ey}
M.~Henningson and K.~Skenderis, ``{Holography and the Weyl anomaly},'' {\em
  Fortsch. Phys.} {\bf 48} (2000)  125--128,
\href{http://arxiv.org/abs/hep-th/9812032}{{\tt arXiv:hep-th/9812032}}.
%%CITATION = HEP-TH/9812032;%%.

\bibitem{Balasubramanian:1999re}
V.~Balasubramanian and P.~Kraus, ``{A stress tensor for anti-de Sitter
  gravity},'' \href{http://dx.doi.org/10.1007/s002200050764}{{\em Commun. Math.
  Phys.} {\bf 208} (1999)  413--428},
\href{http://arxiv.org/abs/hep-th/9902121}{{\tt arXiv:hep-th/9902121}}.
%%CITATION = HEP-TH/9902121;%%.

\bibitem{deHaro:2000xn}
S.~de~Haro, S.~N. Solodukhin, and K.~Skenderis, ``{Holographic reconstruction
  of spacetime and renormalization in the AdS/CFT correspondence},''
  \href{http://dx.doi.org/10.1007/s002200100381}{{\em Commun. Math. Phys.} {\bf
  217} (2001)  595--622},
\href{http://arxiv.org/abs/hep-th/0002230}{{\tt arXiv:hep-th/0002230}}.
%%CITATION = HEP-TH/0002230;%%.

\bibitem{Skenderis:2000in}
K.~Skenderis, ``{Asymptotically anti-de Sitter spacetimes and their stress
  energy tensor},'' \href{http://dx.doi.org/10.1142/S0217751X0100386X}{{\em
  Int. J. Mod. Phys.} {\bf A16} (2001)  740--749},
\href{http://arxiv.org/abs/hep-th/0010138}{{\tt arXiv:hep-th/0010138}}.
%%CITATION = HEP-TH/0010138;%%.

\bibitem{Bianchi:2001de}
M.~Bianchi, D.~Z. Freedman, and K.~Skenderis, ``{How to go with an RG flow},''
  {\em JHEP} {\bf 08} (2001)  041,
\href{http://arxiv.org/abs/hep-th/0105276}{{\tt arXiv:hep-th/0105276}}.
%%CITATION = HEP-TH/0105276;%%.

\bibitem{Bianchi:2001kw}
M.~Bianchi, D.~Z. Freedman, and K.~Skenderis, ``{Holographic
  Renormalization},'' {\em Nucl. Phys.} {\bf B631} (2002)  159--194,
\href{http://arxiv.org/abs/hep-th/0112119}{{\tt arXiv:hep-th/0112119}}.
%%CITATION = HEP-TH/0112119;%%.

\bibitem{Skenderis:2002wp}
K.~Skenderis, ``{Lecture notes on holographic renormalization},''
  \href{http://dx.doi.org/10.1088/0264-9381/19/22/306}{{\em Class. Quant.
  Grav.} {\bf 19} (2002)  5849--5876},
\href{http://arxiv.org/abs/hep-th/0209067}{{\tt arXiv:hep-th/0209067}}.
%%CITATION = HEP-TH/0209067;%%.

\bibitem{Papadimitriou:2004ap}
I.~Papadimitriou and K.~Skenderis, ``{AdS / CFT correspondence and geometry},''
\href{http://arxiv.org/abs/hep-th/0404176}{{\tt arXiv:hep-th/0404176}}.
%%CITATION = HEP-TH/0404176;%%.

\bibitem{Papadimitriou:2004rz}
I.~Papadimitriou and K.~Skenderis, ``{Correlation functions in holographic RG
  flows},'' \href{http://dx.doi.org/10.1088/1126-6708/2004/10/075}{{\em JHEP}
  {\bf 10} (2004)  075},
\href{http://arxiv.org/abs/hep-th/0407071}{{\tt arXiv:hep-th/0407071}}.
%%CITATION = HEP-TH/0407071;%%.

\bibitem{Itzhaki:1998dd}
N.~Itzhaki, J.~M. Maldacena, J.~Sonnenschein, and S.~Yankielowicz,
  ``{Supergravity and the large N limit of theories with sixteen
  supercharges},'' \href{http://dx.doi.org/10.1103/PhysRevD.58.046004}{{\em
  Phys. Rev.} {\bf D58} (1998)  046004},
\href{http://arxiv.org/abs/hep-th/9802042}{{\tt arXiv:hep-th/9802042}}.
%%CITATION = HEP-TH/9802042;%%.

\bibitem{Sekino:1999av}
Y.~Sekino and T.~Yoneya, ``{Generalized AdS-CFT correspondence for matrix
  theory in the large N limit},''
  \href{http://dx.doi.org/10.1016/S0550-3213(99)00793-2}{{\em Nucl. Phys.} {\bf
  B570} (2000)  174--206},
\href{http://arxiv.org/abs/hep-th/9907029}{{\tt arXiv:hep-th/9907029}}.
%%CITATION = HEP-TH/9907029;%%.

\bibitem{Cai:1999xg}
R.-G. Cai and N.~Ohta, ``{Surface counterterms and boundary stress-energy
  tensors for asymptotically non-anti-de Sitter spaces},''
  \href{http://dx.doi.org/10.1103/PhysRevD.62.024006}{{\em Phys. Rev.} {\bf
  D62} (2000)  024006},
\href{http://arxiv.org/abs/hep-th/9912013}{{\tt arXiv:hep-th/9912013}}.
%%CITATION = HEP-TH/9912013;%%.

\bibitem{Sekino:2000mg}
Y.~Sekino, ``{Supercurrents in matrix theory and the generalized AdS/CFT
  correspondence},''
  \href{http://dx.doi.org/10.1016/S0550-3213(01)00126-2}{{\em Nucl. Phys.} {\bf
  B602} (2001)  147--171},
\href{http://arxiv.org/abs/hep-th/0011122}{{\tt arXiv:hep-th/0011122}}.
%%CITATION = HEP-TH/0011122;%%.

\bibitem{Gherghetta:2001iv}
T.~Gherghetta and Y.~Oz, ``{Supergravity, non-conformal field theories and
  brane- worlds},'' \href{http://dx.doi.org/10.1103/PhysRevD.65.046001}{{\em
  Phys. Rev.} {\bf D65} (2002)  046001},
\href{http://arxiv.org/abs/hep-th/0106255}{{\tt arXiv:hep-th/0106255}}.
%%CITATION = HEP-TH/0106255;%%.

\bibitem{Asano:2003xp}
M.~Asano, Y.~Sekino, and T.~Yoneya, ``{PP-wave holography for Dp-brane
  backgrounds},'' \href{http://dx.doi.org/10.1016/j.nuclphysb.2003.11.005}{{\em
  Nucl. Phys.} {\bf B678} (2004)  197--232},
\href{http://arxiv.org/abs/hep-th/0308024}{{\tt arXiv:hep-th/0308024}}.
%%CITATION = HEP-TH/0308024;%%.

\bibitem{Asano:2004vj}
M.~Asano and Y.~Sekino, ``{Large N limit of SYM theories with 16 supercharges
  from superstrings on Dp-brane backgrounds},''
  \href{http://dx.doi.org/10.1016/j.nuclphysb.2004.11.007}{{\em Nucl. Phys.}
  {\bf B705} (2005)  33--59},
\href{http://arxiv.org/abs/hep-th/0405203}{{\tt arXiv:hep-th/0405203}}.
%%CITATION = HEP-TH/0405203;%%.

\bibitem{Wiseman:2008qa}
T.~Wiseman and B.~Withers, ``{Holographic renormalization for coincident
  Dp-branes},'' \href{http://dx.doi.org/10.1088/1126-6708/2008/10/037}{{\em
  JHEP} {\bf 10} (2008)  037},
\href{http://arxiv.org/abs/0807.0755}{{\tt arXiv:0807.0755 [hep-th]}}.
%%CITATION = 0807.0755;%%.

\bibitem{Kanitscheider:2008kd}
I.~Kanitscheider, K.~Skenderis, and M.~Taylor, ``{Precision holography for
  non-conformal branes},''
  \href{http://dx.doi.org/10.1088/1126-6708/2008/09/094}{{\em JHEP} {\bf 09}
  (2008)  094},
\href{http://arxiv.org/abs/0807.3324}{{\tt arXiv:0807.3324 [hep-th]}}.
%%CITATION = 0807.3324;%%.

\bibitem{Kanitscheider:2009as}
I.~Kanitscheider and K.~Skenderis, ``{Universal hydrodynamics of non-conformal
  branes},''
\href{http://arxiv.org/abs/0901.1487}{{\tt arXiv:0901.1487 [hep-th]}}.
%%CITATION = 0901.1487;%%.

\bibitem{Duff:1994fg}
M.~J. Duff, G.~W. Gibbons, and P.~K. Townsend, ``{Macroscopic superstrings as
  interpolating solitons},''
  \href{http://dx.doi.org/10.1016/0370-2693(94)91260-2}{{\em Phys. Lett.} {\bf
  B332} (1994)  321--328},
\href{http://arxiv.org/abs/hep-th/9405124}{{\tt arXiv:hep-th/9405124}}.
%%CITATION = HEP-TH/9405124;%%.

\bibitem{Boonstra:1997dy}
H.~J. Boonstra, B.~Peeters, and K.~Skenderis, ``{Duality and asymptotic
  geometries},'' \href{http://dx.doi.org/10.1016/S0370-2693(97)01008-3}{{\em
  Phys. Lett.} {\bf B411} (1997)  59--67},
\href{http://arxiv.org/abs/hep-th/9706192}{{\tt arXiv:hep-th/9706192}}.
%%CITATION = HEP-TH/9706192;%%.

\bibitem{Boonstra:1998yu}
H.~J. Boonstra, B.~Peeters, and K.~Skenderis, ``{Brane intersections, anti-de
  Sitter spacetimes and dual superconformal theories},''
  \href{http://dx.doi.org/10.1016/S0550-3213(98)00512-4}{{\em Nucl. Phys.} {\bf
  B533} (1998)  127--162},
\href{http://arxiv.org/abs/hep-th/9803231}{{\tt arXiv:hep-th/9803231}}.
%%CITATION = HEP-TH/9803231;%%.

\bibitem{Boonstra:1998mp}
H.~J. Boonstra, K.~Skenderis, and P.~K. Townsend, ``{The domain wall/QFT
  correspondence},'' {\em JHEP} {\bf 01} (1999)  003,
\href{http://arxiv.org/abs/hep-th/9807137}{{\tt arXiv:hep-th/9807137}}.
%%CITATION = HEP-TH/9807137;%%.

\bibitem{Jevicki:1998ub}
A.~Jevicki, Y.~Kazama, and T.~Yoneya, ``{Generalized conformal symmetry in
  D-brane matrix models},''
  \href{http://dx.doi.org/10.1103/PhysRevD.59.066001}{{\em Phys. Rev.} {\bf
  D59} (1999)  066001},
\href{http://arxiv.org/abs/hep-th/9810146}{{\tt arXiv:hep-th/9810146}}.
%%CITATION = HEP-TH/9810146;%%.

\bibitem{Skenderis:1998dq}
K.~Skenderis, ``{Field theory limit of branes and gauged supergravities},''
  {\em Fortsch. Phys.} {\bf 48} (2000)  205--208,
\href{http://arxiv.org/abs/hep-th/9903003}{{\tt arXiv:hep-th/9903003}}.
%%CITATION = HEP-TH/9903003;%%.

\bibitem{Karch:2002sh}
A.~Karch and E.~Katz, ``{Adding flavor to AdS/CFT},'' {\em JHEP} {\bf 06}
  (2002)  043,
\href{http://arxiv.org/abs/hep-th/0205236}{{\tt arXiv:hep-th/0205236}}.
%%CITATION = HEP-TH/0205236;%%.

\bibitem{Karch:2005ms}
A.~Karch, A.~O'Bannon, and K.~Skenderis, ``{Holographic renormalization of
  probe D-branes in AdS/CFT},'' {\em JHEP} {\bf 04} (2006)  015,
\href{http://arxiv.org/abs/hep-th/0512125}{{\tt arXiv:hep-th/0512125}}.
%%CITATION = HEP-TH/0512125;%%.

\bibitem{Skenderis:2002vf}
K.~Skenderis and M.~Taylor, ``{Branes in AdS and pp-wave spacetimes},'' {\em
  JHEP} {\bf 06} (2002)  025,
\href{http://arxiv.org/abs/hep-th/0204054}{{\tt arXiv:hep-th/0204054}}.
%%CITATION = HEP-TH/0204054;%%.

\bibitem{Breitenlohner:1982bm}
P.~Breitenlohner and D.~Z. Freedman, ``{Positive Energy in anti-De Sitter
  Backgrounds and Gauged Extended Supergravity},''
\href{http://dx.doi.org/10.1016/0370-2693(82)90643-8}{{\em Phys. Lett.} {\bf
  B115} (1982)  197}.
%%CITATION = PHLTA,B115,197;%%.

\bibitem{Wang:2003yc}
X.-J. Wang and S.~Hu, ``{Intersecting branes and adding flavors to the
  Maldacena- Nunez background},'' {\em JHEP} {\bf 09} (2003)  017,
\href{http://arxiv.org/abs/hep-th/0307218}{{\tt arXiv:hep-th/0307218}}.
%%CITATION = HEP-TH/0307218;%%.

\bibitem{Kruczenski:2003uq}
M.~Kruczenski, D.~Mateos, R.~C. Myers, and D.~J. Winters, ``{Towards a
  holographic dual of large-N(c) QCD},'' {\em JHEP} {\bf 05} (2004)  041,
\href{http://arxiv.org/abs/hep-th/0311270}{{\tt arXiv:hep-th/0311270}}.
%%CITATION = HEP-TH/0311270;%%.

\bibitem{Sakai:2004cn}
T.~Sakai and S.~Sugimoto, ``{Low energy hadron physics in holographic QCD},''
  \href{http://dx.doi.org/10.1143/PTP.113.843}{{\em Prog. Theor. Phys.} {\bf
  113} (2005)  843--882},
\href{http://arxiv.org/abs/hep-th/0412141}{{\tt arXiv:hep-th/0412141}}.
%%CITATION = HEP-TH/0412141;%%.

\bibitem{Sakai:2005yt}
T.~Sakai and S.~Sugimoto, ``{More on a holographic dual of QCD},''
  \href{http://dx.doi.org/10.1143/PTP.114.1083}{{\em Prog. Theor. Phys.} {\bf
  114} (2005)  1083--1118},
\href{http://arxiv.org/abs/hep-th/0507073}{{\tt arXiv:hep-th/0507073}}.
%%CITATION = HEP-TH/0507073;%%.

\bibitem{Burrington:2009fm}
B.~A. Burrington and J.~Sonnenschein, ``{Holographic Dual of QCD from Black D5
  Branes},''
\href{http://arxiv.org/abs/0903.0628}{{\tt arXiv:0903.0628 [hep-th]}}.
%%CITATION = 0903.0628;%%.

\bibitem{Mezincescu:1984ev}
L.~Mezincescu and P.~K. Townsend, ``{Stability at a Local Maximum in Higher
  Dimensional Anti-de Sitter Space and Applications to Supergravity},''
\href{http://dx.doi.org/10.1016/0003-4916(85)90150-2}{{\em Ann. Phys.} {\bf
  160} (1985)  406}.
%%CITATION = APNYA,160,406;%%.

\end{thebibliography}\endgroup

\end{document}